\documentclass[12pt]{article}
\usepackage[utf8]{inputenc}

\usepackage{amsfonts,amsthm,amsmath,amssymb,upgreek,bm}
\usepackage[hidelinks]{hyperref}
\usepackage[paper=letterpaper,margin=.85in]{geometry}
\usepackage{graphicx}
\usepackage{units}
\usepackage{float}
\usepackage[export]{adjustbox}
\usepackage{bbold}
\usepackage{xcolor}
\usepackage{braket}
\usepackage{newunicodechar}
\newunicodechar{，}{,}
\usepackage{braket}
\usepackage[shortlabels]{enumitem}
\usepackage{dsfont}
\usepackage[mathscr]{euscript}
\usepackage[normalem]{ulem}
\usepackage{physics}
\usepackage{tikz-cd}
\usepackage[autostyle]{csquotes}
\usepackage{babel}

\newcommand{\be}{\begin{equation}}
\newcommand{\ee}{\end{equation}}

\numberwithin{equation}{section}

\def\Tr{\text{Tr}}

\usepackage{diagbox} 
\usepackage{multirow} 
\usepackage{xcolor}
\definecolor{dgreen}{rgb}{0.0, 0.5, 0.0}
\definecolor{dred}{rgb}{0.5, 0.0 , 0.0}
\usepackage{slashed}
 
\usepackage[compress]{cite}

\usepackage{braket}

\usepackage{physics}

	\usepackage{tikz}
	\usepackage{makecell}

\usepackage{amsthm} 

\def\ie{\begin{equation}\begin{aligned}}
\def\fe{\end{aligned}\end{equation}}

\newcommand{\RN}[1]{%
  \textup{\uppercase\expandafter{\romannumeral#1}}%
}

\usepackage{microtype}
\begin{document}

\nocite{*}

\thispagestyle{empty}

\vspace*{2.5cm}
\begin{center}

{\bf {\LARGE Adapted Fröhlich–Pizzo Model for Horizon-Induced Decoherence }}

\begin{center}

\vspace{1cm}

{\bf  Bik-Soon Sia$^{1 \dagger}$}    \\
  \bigskip \rm

\bigskip 
${}^1$ Independent Researcher

\bigskip
\rmfamily ${}^{\dagger}$\textit{Email}: \texttt{b.s.sia1106@gmail.com} 
\rm
  \end{center}

\vspace{2cm}
{\bf Abstract}
\end{center}
\begin{quotation}
\noindent

The Fröhlich–Pizzo (FP) model is a quantum-mechanical model originally developed to describe the interaction between an atom and the electromagnetic field in the nonrelativistic limit $c\to\infty$. We show that an adaptation of the FP model based on the local algebras of a $U(1)$ chiral current captures essential features of the horizon-induced decoherence mechanism identified by Danielson, Satishchandran, and Wald (DSW) in the limit of large holding time $T$. Our analysis suggests that the effective dynamics of the system undergoing decoherence in the DSW setting retains non-Markovian features. Our approach also offers a glimpse into the possible role of quantum events in  the dynamics underlying horizon-induced decoherence.

\end{quotation}

\setcounter{page}{0}
\setcounter{tocdepth}{2}
\setcounter{footnote}{0}
\newpage

\setcounter{page}{2}

\tableofcontents

\section{Introduction}\label{sec:introduction}

One of the longstanding conceptual difficulties in quantum
mechanics concerns the status of measurement and the
occurrence of quantum events. While the unitary evolution
of quantum states provides a remarkably successful
description of microscopic phenomena, the appearance of
definite measurement outcomes is conventionally described
by an additional rule of state reduction. The distinction
between unitary evolution and state collapse raises a
fundamental question: can the occurrence of quantum events
be incorporated into a unified dynamical description,
without treating measurement as an externally prescribed
intervention?

A profound proposal in this direction is the
Events--Trees--Histories (ETH) approach to quantum
mechanics,\footnote{The idea that quantum events should
play a central role in the description of physical
phenomena was already advocated by Rudolf Haag, whose
work provided an important conceptual precursor to
the ETH approach
\cite{Haag:1990ez,Haag:2013gna,Haag:1996dw}.}
developed by Fr\"ohlich and collaborators
\cite{Frohlich:2021gaf,Froehlich:2019qnd,
Frohlich:2026epd,Frohlich:2024kyo,Frohlich:2019bsc}.
Its central idea is that the occurrence of quantum
events should be understood as an intrinsic aspect
of quantum dynamics. A central mathematical ingredient
of this approach is the Principle of Diminishing
Potentialities (PDP), according to which the von Neumann algebras associated with future potentialities form a strictly decreasing family as time progresses.,
\begin{align}
    \mathcal E_{\geq t'}
    \subsetneq
    \mathcal E_{\geq t},
    \qquad t'>t.
\end{align}
This structure captures the gradual disappearance of potentialities with the passage of time. More importantly, the ETH approach provides
an algebraic criterion for determining the occurrence
of actual quantum events through the center of the
centralizer of the state restricted to the algebra
of future potentialities. Once an event occurs,
one of its possible outcomes is realized according
to Born's rule, and the corresponding state
undergoes reduction. Repeated occurrences of such
events give rise to a stochastic branching dynamics,
in which a single realized history emerges from a
tree of possible quantum histories. In this way,
the ETH approach places the occurrence of quantum
events and the intrinsic stochasticity of quantum
dynamics at the heart of its theoretical description.

A particularly instructive realization of these ideas is provided by the Fröhlich–Pizzo (FP) model \cite{Frohlich:2021gaf,Froehlich2019}. Originally motivated by the interaction between an atom and the electromagnetic radiation field in the nonrelativistic limit ($c\to\infty$), the FP model provides a concrete dynamical setting in which the central ingredients of the ETH approach can be studied explicitly. The model consists of a finite-dimensional ``atomic'' system coupled to an infinite collection of environmental degrees of freedom, referred to as $R$-fields. Its dynamics proceeds in discrete time steps, with the interaction at each step involving the atomic system and the corresponding $R$-field degrees of freedom. The resulting family of algebras of future potentialities is strictly decreasing as time progresses, thereby providing a concrete realization of PDP. What makes the FP model particularly interesting is that it allows the occurrence of quantum events to be analyzed explicitly within its algebraic dynamics. Owing to the special structure of the interaction and the choice of reference state for the $R$-fields, the state restricted to the algebra of future potentialities admits a factorized description in terms of an updated atomic density matrix and an independent reference state of the remaining $R$ fields. This property makes it possible to determine the relevant center of the centralizer from the spectral projections of the effective atomic density matrix. Together with the state-reduction rule prescribed by the ETH approach, this leads to a recursive stochastic dynamics in which the occurrence of quantum events and the subsequent evolution of the physical state can be followed explicitly. Although the FP model was originally motivated by a particular atom–radiation interaction, its mathematical construction admits a broader interpretation. The finite-dimensional ``atomic'' system may represent a more general quantum system, while the $R$-fields may be interpreted as environmental degrees of freedom with which it interacts. This suggests the possibility of adapting the FP framework to other physical settings, provided that the essential algebraic  structures can be appropriately realized. 

Meanwhile, an important development in the study
of quantum decoherence in curved spacetime was
made by Danielson, Satishchandran, and Wald (DSW),
who showed that the presence of a black hole
can induce decoherence of nearby quantum
superpositions \cite{Danielson:2022tdw}.
Their analysis further established that this
phenomenon is not restricted to black holes,
but arises more generally in spacetimes
possessing Killing horizons
\cite{Danielson:2022sga}.
In the DSW setting, a charged particle is
prepared in a quantum superposition of two
spatially separated configurations and held
stationary near a Killing horizon. The
electromagnetic field sourced by the particle
extends to the horizon, where the two
configurations leave distinguishable imprints.
Information about the position of the charged
particle consequently becomes encoded in
the electromagnetic degrees of freedom
associated with the horizon, leading to
a gradual suppression of coherence between
the two branches. The corresponding decoherence
can be described in terms of the overlap
between the electromagnetic field states
associated with the alternative particle
configurations. This discovery is particularly
striking because the particle may remain
stationary during the holding period, yet
its quantum coherence continues to decay.
These developments have established
horizon-induced decoherence as a significant
phenomenon at the interface of quantum
mechanics, quantum field theory, and
gravitational physics.

The purpose of the present paper is to explore a possible
connection between the ETH approach to quantum mechanics
and the phenomenon of horizon-induced decoherence.
Although these two developments arise from rather
different physical motivations, they share an important
underlying theme: the interaction of a quantum system
with environmental degrees of freedom and the resulting
loss of quantum coherence. In the DSW setting, the
horizon provides a natural environment in which
information about the alternative configurations
of the charged particle becomes encoded. From the
perspective of the ETH approach, it is then natural
to ask whether the dynamics underlying this
decoherence mechanism can be understood within
a framework that also incorporates the possible
occurrence of quantum events.

A particularly appealing feature of the FP model
is that its underlying dynamical structure is not
restricted to the original ``atom''-radiation system.
This suggests that an appropriate adaptation of
the model may provide a useful description of
horizon-induced decoherence. In making such an
adaptation, we identify the two configurations
of the charged particle with the states of the
two-level ``atomic'' system, while the degrees
of freedom associated with the horizon play the
role of the $R$-fields. The interaction between
the two systems is then understood in terms of
the branch-dependent electromagnetic information
imprinted on the horizon. In this way, the
physical mechanism responsible for horizon-induced
decoherence can be described using the basic
dynamical ingredients of the FP construction. However, a direct application of the original
discrete FP model to the horizon is not entirely
straightforward. The horizon degrees of freedom
are described by a continuum quantum field theory,
rather than by an infinite collection of independent
finite-dimensional subsystems. Moreover, the
underlying algebraic structure must be compatible
with the causal structure of a Killing horizon.
To address these issues, we propose an adapted
FP model based on the local von Neumann algebras
of a $U(1)$ chiral current. The use of this model
allows us to retain essential features of the
original FP construction while incorporating
the continuum structure of quantum fields on
the horizon.

\par An important observation is that the natural
time parameter for our construction is the
logarithm of the affine null coordinate along
the horizon. With this choice, the future
horizon algebras form a decreasing family
under the advancement of the horizon cut,
providing a realization of the PDP. Furthermore,
the vacuum state restricted to the future
horizon algebra is a KMS state with respect
to the corresponding dilation flow. These
properties make the chiral-current model
particularly suitable for investigating
the relationship between horizon-induced
decoherence and the algebraic structure
underlying the ETH approach.

\par Within this adapted construction, we show
that the off-diagonal Gram coefficients which  characterizes the distinguishability of the horizon field states associated with different configurations of the charged particle during the interaction period
exhibit exponential suppression in the
large holding-time limit. Consequently,
the reduced state of the charged particle
reproduces the characteristic asymptotic
decoherence behavior identified by DSW.  However, an important difference shows up when we compare our adapted model with the original discrete FP model. In the latter, the Gram matrix can be completely absorbed into the effective state of the ``atomic'' system, leaving the remaining $R$-fields in an independent reference state. In our continuum construction,  we find that nontrivial vacuum correlations between the interaction region and the future horizon prevent such an exact factorization. These residual correlations, which are not explicitly examined in the original DSW analysis, suggest possible non-Markovian features in the dynamics underlying horizon-induced decoherence. In particular, although the reduced dynamics of the charged particle admits a simple effective description in the large holding-time limit, the full particle--horizon state retains correlations with the earlier interaction that cannot be inferred from the asymptotic reduced decoherence law alone.  This failure of factorization may also have implications for the occurrence of quantum events within the ETH approach. In the original discrete FP model, the event structure can be determined from the spectral projections of the effective atomic density matrix, owing to the factorization of the state. In our adapted model, however, this simplification is no longer directly available, since the factorization does not hold and the center of the centralizer must be determined from the full interacting state on the combined particle--horizon algebra. This suggests that the horizon degrees of freedom may play an important role not only in inducing decoherence, but also in determining the structure of quantum events.

The remainder of this paper is organized as follows.
In Section 2, we provide an overview of the
Events, Trees and Histories (ETH) approach to
quantum mechanics, with particular emphasis on
the Principle of Diminishing Potentialities (PDP)
and the algebraic criteria for the occurrence of
quantum events. In Section 3, we review the
Fr\"ohlich--Pizzo (FP) model and explain how
the essential ingredients of the ETH approach
are realized within this concrete dynamical
framework, including the emergence of stochastic
dynamics through successive quantum events.
In Section 4, we revisit the horizon-induced
decoherence mechanism identified by Danielson,
Satishchandran and Wald (DSW), focusing on
the physical setup and the calculation of
decoherence induced by the distinguishability
of electromagnetic field states on the horizon. In Section 5, we propose an adaptation of the
FP model based on the local algebras of a
$U(1)$ chiral current. We examine its algebraic
structure, derive the characteristic asymptotic
decoherence behavior, and investigate the residual
correlations between the interaction region and
the future horizon. We also discuss the possible
implications of these correlations for the
underlying dynamics and the occurrence of
quantum events. Finally, in Section 6, we
summarize our findings and discuss several
open questions and reflection for
future investigation.

\section{A Brief Overview of ETH Approach} 
In this section, we want to give a brief overview of Events-Trees-Histories approach to quantum mechanics  \cite{Frohlich:2021gaf,Froehlich:2019qnd}.  ETH approach proposes a completion of quantum mechanics in which the occurrence of quantum events becomes an intrinsic part of the dynamical description, rather than being introduced through an externally prescribed measurement process. Its central insight is that the unitary evolution of observables and the stochastic evolution of states are not necessarily competing descriptions, but may be understood as complementary aspects of quantum dynamics. We briefly review the essential mathematical structure of this approach, with particular emphasis on the Principle of Diminishing Potentialities (PDP), the algebraic characterization of quantum events, and the resulting stochastic branching of quantum histories.

\par The fundamental object in the ETH approach is an
\textit{isolated open system} $S$. Such a system is effectively
isolated from its complement during the timescale of
interest, but remains open in the sense that it may emit
modes which subsequently escape its observational reach.
Although these modes can no longer be detected through
measurements performed on $S$, they may remain entangled
with its remaining degrees of freedom. Now, we
let $\mathcal{H}_S$ denote the Hilbert space of the system,
and let $\mathcal{A}_{\geq t}$ be the von Neumann algebra
generated by all observables that can, in principle, be
measured at time $t$ or later. The family $\left\{
    \mathcal{A}_{\geq t}
    \right\}_{t\in\mathbb{R}}$
forms a decreasing family, or more precisely a
\textit{co-filtration}, of von Neumann algebras.
The Principle of Diminishing Potentialities (PDP)
asserts that $\mathcal{A}_{\geq t}
    \subsetneq
    \mathcal{A}_{\geq t'},
    \: t>t'$. The strictness of this inclusion is crucial.
It expresses the progressive loss of observational
possibilities as time advances: certain observables
accessible at an earlier time are no longer accessible
at later times. Thus, PDP introduces a fundamental
temporal asymmetry into the algebraic structure of
the theory, even when the underlying evolution of
observables is described by a one-parameter unitary
group.  We also introduce the full algebra of potentialities $ \mathcal{A}
    =
    \overline{
    \bigcup_{t\in\mathbb{R}}
    \mathcal{A}_{\geq t}
    }^{\|\cdot\|}$
together with a state $\omega$ on $\mathcal{A}$. In the Heisenberg picture, the state is fixed,
but its restriction to the algebra of future
potentialities, $\omega_t
    =
    \omega|_{\mathcal{A}_{\geq t}}$
changes as the accessible algebra shrinks.
In particular, a state that is pure on the full
algebra may appear mixed when restricted to
$\mathcal{A}_{\geq t}$. This observation provides
the starting point for the algebraic description
of quantum events.

\par A \textit{potential event} at time $t$ is represented
by a partition of unity consisting of mutually
orthogonal projections in $\mathcal{A}_{\geq t}$,
\begin{align}
    \left\{\pi_\xi\right\}_{\xi\in X},
    \qquad
    \pi_\xi\pi_\eta
    =
    \delta_{\xi\eta}\pi_\xi,
    \qquad
    \sum_{\xi\in X}\pi_\xi
    =
    \mathbb{1}.
\end{align}
where $X$ here is a countable set. The central question is how to determine which
of these potential events actually occurs,
without arbitrarily choosing a measurement
observable or a preferred basis.
The first step is to identify partitions of unity
for which the restricted state admits an incoherent
decomposition,
\begin{align}
    \omega_t(A)
    =
    \sum_{\xi\in X}
    \omega_t(\pi_\xi A\pi_\xi),
    \qquad
    \forall A\in\mathcal{A}_{\geq t}.
    \label{eq:ETH_decomposition}
\end{align}
Such decompositions are naturally related to the
\textit{centralizer} of the state $\omega_t$,
\begin{align}
    \mathcal{C}_{\omega_t}
    (\mathcal{A}_{\geq t})
    =
    \left\{
    Y\in\mathcal{A}_{\geq t}
    \,\middle|\,
    \omega_t([Y,A])=0,\quad
    \forall A\in\mathcal{A}_{\geq t}
    \right\}.
    \label{eq:ETH_centralizer}
\end{align}
The spectral projections of any self-adjoint
element of the centralizer satisfy
(\ref{eq:ETH_decomposition}).
However, the centralizer may contain noncommuting
observables, whose spectral projections define
mutually incompatible decompositions.
Consequently, the centralizer alone does not
necessarily single out a unique event structure. The ETH approach resolves this ambiguity by
passing to the \textit{center of the centralizer}
\begin{align}
    Z_{\omega_t}
    (\mathcal{A}_{\geq t})
    =
    \mathcal{C}_{\omega_t}
    (\mathcal{A}_{\geq t})
    \cap
    \mathcal{C}_{\omega_t}
    (\mathcal{A}_{\geq t})'
    \label{eq:ETH_center}
\end{align}
This is an abelian von Neumann algebra, whose
spectral decomposition specifies the distinguished
family of mutually compatible event projections.
In the ETH prescription, a nontrivial actual event
is associated with a nontrivial center of the
centralizer, containing projections with
probabilities strictly between zero and one.
The event structure is therefore determined
jointly by the algebra of future potentialities
and the state of the system, rather than by an
external observer. The center of the centralizer plays a decisive role
in determining which potential events become actual
events. More precisely, when
$Z_{\omega_t}(\mathcal{A}_{\geq t})$
is nontrivial, its spectral projections specify
a distinguished partition of unity
$\{\pi_\xi\}_{\xi \in X_{\omega_t}}$,
which is identified with the actual event occurring
at time $ \geq t$.  An actual event represented by
the partition of unity
$\{\pi_\xi\}_{\xi\in X_{\omega_t}}$
is said to occur at time $t$ if
$\{\pi_\xi\}_{\xi\in X_{\omega_t}}
    \subset \mathcal{A}_{\geq t}$ but $\{\pi_\xi\}_{\xi\in X_{\omega_t}}
    \not\subset \mathcal{A}_{\geq t'},
    \: \forall\,t'>t$.

\par Once an actual event occurs, the ETH approach
supplements the unitary evolution of observables
with a stochastic law governing the state.
Suppose that the event is represented by
projections $\{\pi_\xi\}_{\xi\in X}$.
One outcome $\xi_*$ is realized with probability
given by Born's rule
\begin{align}
    \mathbb{P}(\xi_*,t)
    =
    \omega_t(\pi_{\xi_*}),
\end{align}
and the corresponding state is updated according
to the Collapse Postulate (CP),
\begin{align}
    \omega_{t_+,\xi_*}(A)
    =
    \frac{
    \omega_{t_+}(\pi_{\xi_*}A\pi_{\xi_*})
    }{
    \omega_{t_+}(\pi_{\xi_*})
    }.
    \label{eq:ETH_collapse}
\end{align}
Here $t_+$ denotes a time immediately following
the occurrence of the event, and the restriction
to the corresponding future algebra is understood.

The resulting evolution is a nonlinear stochastic
branching process. At each actual event, the system
selects one of several possible continuations,
with probabilities determined by the state immediately
preceding the event. Repeated applications of this
prescription generate a tree of possible quantum
histories, while a single sequence of realized
outcomes constitutes the actual history of the
system. This distinction is conceptually important.
Unlike the Everett interpretation, in which
all branches are retained as components of a
universal quantum state \cite{RevModPhys.29.454}, the ETH approach
interprets the branching structure stochastically,
with only one history being realized.
Quantum events are thus not merely records of
decoherence, but constitute the successive
actualities through which the physical history
of the system unfolds. Moreover, if one averages over all possible histories, the effective/averaged evolution is the standard linear quantum dynamics governed by the Schrödinger equation.

\par For a more detailed discussion of the physical mechanisms underlying the Principle of Diminishing Potentialities (PDP), the extension of the ETH approach to relativistic quantum field theory, and its connections to large $N$ algebras in holography,
see \cite{Sia:2025rux,Froehlich:2019qnd,Frohlich:2019bsc}.

 \section{Fr\"ohlich-Pizzo Model}
In this section, we examine the Fr\"ohlich--Pizzo (FP)
model \cite{Frohlich:2021gaf,Froehlich2019},
originally developed to describe the interaction
between an atom and the electromagnetic field in
the nonrelativistic limit, and use it to illustrate
the Events, Trees and Histories (ETH) approach to
quantum mechanics. The FP model provides a
conceptually rich setting that brings to light
several fundamental aspects of quantum dynamics.
We believe that its significance extends well
beyond its original physical motivation, as its
underlying mathematical structure is not
intrinsically tied to a particular physical
interpretation. By appropriately reinterpreting
its constituent systems and their interactions,
the FP framework may be adapted to other
physical settings. In particular, as we shall demonstrate in the
following sections, an appropriate adaptation
of the FP model captures essential features
of Killing-horizon-induced decoherence,
providing a useful setting in which to
investigate the quantum dynamics underlying
this phenomenon. This broader applicability
motivates us to adopt a more general terminology
throughout our discussion. We shall refer to
the finite-dimensional subsystem as the
``atomic'' system, with quotation marks
emphasizing its abstract role, and use the
term ``$R$-fields'' for the environmental
degrees of freedom, without identifying
them specifically with the electromagnetic
radiation field.
\subsection{Setup and Features}

We first describe the setup for the $R$-fields. Time is taken to be
discrete,
\[
    \tau\in\mathbb Z=\{\ldots,-2,-1,0,1,2,\ldots\},
\]
and we choose $\tau=0$ as a reference time. Associated with each discrete
time $\tau$ is a finite-dimensional Hilbert space
\[
    \mathcal H_\tau\simeq\mathbb C^N,
\]
together with the local algebra of observables
\[
    \mathcal B_\tau
    =
    \mathcal B(\mathcal H_\tau)
    \simeq
    M_N(\mathbb C).
\]

We let $\{\varphi_0,\varphi_1,\ldots,\varphi_{N-1}\}$ be an
orthonormal basis of $\mathcal H_\tau$, and choose $\varphi_0$ as a
distinguished reference vector. Formally, one may consider the complete
infinite tensor product of the spaces $\mathcal H_\tau$. Such a space is,
however, generally non-separable and contains mutually inequivalent sectors
corresponding to macroscopically distinct infinite configurations; it is
therefore unnecessarily large for the physical Hilbert space considered
here.

Instead, we select the sector generated by finite excitations above the
reference configuration. We define the linear space
\begin{equation}
    \mathcal D
    :=
    \operatorname{span}
    \left\{
        \Phi_{\vec k}
        =
        \bigotimes_{\tau\in\mathbb Z}\varphi_{k(\tau)}
        \;\middle|\;
        k(\tau)\neq0
        \text{ for only finitely many }\tau
    \right\},
    \label{eq:Dfinite}
\end{equation}
where
\[
    \vec k
    =
    \{\ldots,k(-2),k(-1),k(0),k(1),k(2),\ldots\}.
\]
The inner product on elementary tensors is defined by
\begin{equation}
    \langle\Phi_{\vec k},\Phi_{\vec k'}\rangle
    =
    \prod_{\tau\in\mathbb Z}
    \left\langle
        \varphi_{k(\tau)},
        \varphi_{k'(\tau)}
    \right\rangle .
    \label{eq:innerproduct}
\end{equation}
Since $k(\tau)$ and $k'(\tau)$ are nonzero at only finitely many
sites, all but finitely many factors in the product are equal to one,
and hence the inner product is well defined. The physical Hilbert space is
then defined as the completion
\begin{equation}
    \mathcal H_\infty
    :=
    \overline{\mathcal D}^{\,\|\cdot\|}.
    \label{eq:Hinf}
\end{equation}
Thus $\mathcal H_\infty$ is the incomplete infinite tensor product
relative to the reference sequence
\[
    (\ldots,\varphi_0,\varphi_0,\varphi_0,\ldots).
\]
The distinguished reference vector is
\begin{equation}
    \Phi_{\vec 0}
    :=
    \bigotimes_{\tau\in\mathbb Z}\varphi_0,
    \label{eq:vacuum}
\end{equation}
which we interpret as the configuration with no excitation. For a finite time interval $[\tau_1,\tau_2]$, we define the local
time-band algebra
\begin{equation}
    \mathcal B_{[\tau_1,\tau_2]}
    := \bigotimes_{\tau = \tau_1}^{\tau_2}\mathcal{B}_{\geq \tau} \simeq
    \bigotimes_{\tau=\tau_1}^{\tau_2}M_N(\mathbb C),
    \label{eq:bandalgebra}
\end{equation}
where the algebra is understood to act trivially outside the
interval $[\tau_1,\tau_2]$. The algebra of local observables is therefore
\begin{equation}
    \mathcal B_{\mathrm{loc}}
    :=
    \bigcup_{\substack{
        \tau_1\leq\tau_2\\
        \tau_1,\tau_2\in\mathbb Z
    }}
    \mathcal B_{[\tau_1,\tau_2]},
    \label{eq:Bloc}
\end{equation}
and the full quasi-local $C^*$-algebra is its norm completion,
\begin{equation}
    \mathcal B
    :=
    \overline{\mathcal B_{\mathrm{loc}}}^{\,\|\cdot\|}.
    \label{eq:Bquasilocal}
\end{equation}

The reference configuration $\Phi_{\vec 0}$ determines a state
$\omega_{\vec 0}$ on $\mathcal B$ by
\begin{equation}
    \omega_{\vec 0}(b)
    :=
    \left\langle
        \Phi_{\vec 0},
        b\Phi_{\vec 0}
    \right\rangle,
    \qquad b\in\mathcal B.
    \label{eq:omega0}
\end{equation}
Equivalently, $\omega_{\vec 0}$ is the infinite product of the one-site
vector states $\omega_{0,\tau}(\cdot)
    =
    \langle\varphi_0,\cdot\varphi_0\rangle$.
Performing the GNS construction for the pair
$(\mathcal B,\omega_{\vec 0})$ gives $ (\mathcal H_{\vec 0},
      \pi_{\vec 0},
      \Omega_{\vec 0})$.
Since local observables acting on $\Phi_{\vec 0}$ generate precisely the
finite-excitation vectors,
\[
    \mathcal D
    =
    \mathcal B_{\mathrm{loc}}\Phi_{\vec 0},
\]
the resulting GNS Hilbert space is naturally unitarily equivalent to the
Hilbert space constructed above:
\begin{equation}
    \mathcal H_{\vec 0}
    \simeq
    \mathcal H_\infty.
    \label{eq:GNSHinf}
\end{equation}

Moreover, $\omega_{\vec 0}$ is a pure state of the quasi-local
$C^*$-algebra $\mathcal B$. By the purity--irreducibility theorem for
GNS representations, we have $\pi_{\vec 0}(\mathcal B)'
    =
    \mathbb C\,1$.
Consequently,
\begin{equation}
    \pi_{\vec 0}(\mathcal B)''
    =
    \mathcal B(\mathcal H_{\vec 0})
    \simeq
    \mathcal B(\mathcal H_\infty),
    \label{eq:typeI}
\end{equation}
and the von Neumann algebra generated in the reference representation is
therefore a Type $I_\infty$ factor.

More generally, let $h$ be a one-site Hamiltonian on $\mathcal H_\tau$,
and we may consider the Gibbs state
\begin{equation}
    \omega_{\beta,\tau}(a)
    :=
    \frac{
        \operatorname{Tr}(e^{-\beta h}a)
    }{
        \operatorname{Tr}(e^{-\beta h})
    },
    \qquad
    a\in M_N(\mathbb C).
    \label{eq:GibbsLocal}
\end{equation}
The corresponding stationary product state on the quasi-local algebra is
defined by
\begin{equation}
    \omega_\beta
    :=
    \bigotimes_{\tau\in\mathbb Z}
    \omega_{\beta,\tau},
    \label{eq:GibbsProduct}
\end{equation}
meaning that for every local tensor
$b=b_{\tau_1}\otimes\cdots\otimes b_{\tau_n}$,
\begin{equation}
    \omega_\beta(b)
    =
    \prod_{j=1}^n
    \omega_{\beta,\tau_j}(b_{\tau_j}),
    \label{eq:GibbsProductExplicit}
\end{equation}
with extension to $\mathcal B$ by norm continuity. The GNS representation associated with $\omega_\beta$ will be denoted by $(\mathcal H_\beta,\pi_\beta,\Omega_\beta)$
and its von Neumann algebra by $\mathcal M_\beta
    :=
    \pi_\beta(\mathcal B)''$.
For $0<\beta<\infty$, if the one-site Gibbs state is faithful and
nontracial, it is well known that the stationary infinite product construction gives a factor
of Type III; its precise Connes type depends on the spectral ratios of the
one-site Gibbs density matrix.
Furthermore, two limiting cases are also known to be  instructive. In the
infinite-temperature limit $\beta\rightarrow0$,
\[
    \omega_{\beta,\tau}(a)
    \longrightarrow
    \frac{1}{N}\operatorname{Tr}(a).
\]
The resulting infinite product state is tracial, and its GNS von Neumann
algebra is the hyperfinite Type $II_1$ factor. On the other hand, suppose that $h$ has a unique ground state
$\varphi_0$. Then
\[
    \frac{e^{-\beta h}}
         {\operatorname{Tr}(e^{-\beta h})}
    \xrightarrow[\beta\to\infty]{}
    |\varphi_0\rangle\langle\varphi_0|.
\]
Consequently, we have $\omega_\beta
    \xrightarrow[\beta\to\infty]{}
    \omega_{\vec 0}$ on every local observable, and the corresponding zero-temperature GNS
von Neumann algebra reduces to the Type $I_\infty$ factor
\eqref{eq:typeI}.

For the convenience of the discussion in this section, we will mainly take
the state of our $R$-fields system to be a vector
$\Phi_{\vec{k}}\in\mathcal H_\infty$. Another important feature of the present setup is that it provides a
natural realization of the PDP. We define the von Neumann algebra
$\mathcal B_{\geq\tau}$, representing all potential events that may occur
at times $\geq\tau$, by
\begin{equation}
    \mathcal B_{\geq\tau}
    =
    \overline{
    \bigvee_{\substack{
        \tau'\geq\tau\\
        \tau'\,\mathrm{finite}
    }}
    \mathcal B_{[\tau,\tau']}
    }^{\,w},
    \label{eq:futurealgebra}
\end{equation}
where the closure is taken in the weak operator topology in the
representation on $\mathcal H_\infty$. It follows immediately from the
definition that, for $\tau'>\tau$,
\begin{equation}
    \mathcal B_{\geq\tau'}
    \subseteq
    \mathcal B_{\geq\tau}
    \subseteq
    \mathcal B(\mathcal H_\infty).
\end{equation}

In fact, the tensor-product structure allows us to say more. For
$\tau'>\tau$, the degrees of freedom associated with the finite time band
$[\tau,\tau'-1]$ commute with all observables localized at times
$\geq\tau'$. Consequently, the relative commutant of
$\mathcal B_{\geq\tau'}$ inside $\mathcal B_{\geq\tau}$ is
\begin{equation}
    \mathcal B_{\geq\tau'}'
    \cap
    \mathcal B_{\geq\tau}
    =
    \mathcal B_{[\tau,\tau'-1]}.
    \label{eq:relativecommutant}
\end{equation}
Since $\mathcal B_{[\tau,\tau'-1]}$ contains nontrivial observables
whenever $\tau'>\tau$, the inclusion above is strict: 
\begin{equation}
    \mathcal B_{\geq\tau'}
    \subsetneq
    \mathcal B_{\geq\tau},
    \qquad
    \tau'>\tau.
\end{equation}
Thus the family 
$\{\mathcal B_{\geq\tau}\}_{\tau\in\mathbb Z}$
forms a strictly decreasing filtration of von Neumann algebras, i.e., PDP holds \footnote{In the original physical motivation of
the FP model, PDP holds due to Huygens principle \cite{Buchholz:2013bia} \cite{Frohlich:2019bsc}, which implies that in even-dimensional spacetime, the algebra of observables
associated with a later future light cone is
strictly contained in that associated with an
earlier future light cone. In the non-relativistic limit $c \rightarrow \infty$, the light cones open up and thus the future algebra is strctly contained in the past algebra. }.

The discrete time structure introduced above also admits a natural
bilateral time-shift operation. Since $\tau\in\mathbb Z$, the shift is
invertible. We first introduce a unitary shift operator $S$ on
$\mathcal H_\infty$. On the finite-excitation basis vectors, it is defined
by
\begin{equation}
    S\Phi_{\vec{k}}
    := \Phi_{\sigma^{-1}(\vec{k})}=\bigotimes_{\tau\in\mathbb Z}\varphi_{k(\tau-1)}, 
    \quad  S^{-1}\Phi_{\vec{k}}
    := \Phi_{\sigma (\Vec{k})} =
    \bigotimes_{\tau\in\mathbb Z}
    \varphi_{k(\tau+1)}.
    \label{eq:Sshift and inverse}
\end{equation}
Thus $S$ translates every excitation one unit towards the future and its inverse $S^{-1}$ translates every excitation one unit towards the past. In particular, 
this unitary map $S$ also induces a one-step time-translation $*$-endomorphism
$\Theta$ on the algebra $\mathcal{B}_{\geq \tau}$
\footnote{If we view $\Theta$ as a map from $\mathcal{B}_{\geq \tau}$ to itself, $\Theta$ is not an onto map. So, its inverse does not exist.} 
\begin{equation}
        \Theta(E)
        =
        SE S^{-1} \subseteq \mathcal{B}_{\geq \tau +1}, \quad E \in \mathcal{B}_{\geq \tau}
    \label{eq:Theta}
\end{equation}
We will also use the symbol $\phi_{\vec{k}}$ for the corresponding vector state
\begin{align}
    \phi_{\Vec{k}}(\cdot)= \langle \Phi_{\Vec{k}} \vert \cdot \vert \Phi_{\Vec{k}} \rangle
\end{align}
and we note that
\begin{align}
    \phi_{\Vec{k}} (SES^{-1})= \langle S^{-1}\Phi_{\Vec{k}}  \vert  E \vert S^{-1} \Phi_{\vec{k}} \rangle = \langle\Phi_{\sigma (\vec{k})} \vert E \vert \Phi_{\sigma(\vec{k})} \rangle = \phi_{\sigma(\vec{k})} (E)
\end{align}
The shift map $S$ plays the role of the unitary that ``generates the free evolution of the $R$ fields'' before the $R$ fields start to interact with the atom.

\par We now introduce the finite-dimensional quantum system that will play the
role of the ``atom.'' Its Hilbert space is taken to be
\begin{equation}
    \mathcal H_A \simeq \mathbb C^M ,
    \label{eq:atomHilbert}
\end{equation}
and its algebra of observables is
\begin{equation}
    \mathcal A
    :=
    \mathcal B(\mathcal H_A)
    \simeq
    M_M(\mathbb C).
    \label{eq:atomAlgebra}
\end{equation}

In the absence of interactions with the $R$-fields, since time is discrete in the present model, the time evolution of the `` atom'' over each time step is given by an unitary $V \in \mathcal{U}(\mathcal{H}_A)$.  
 Thus, after $n\in\mathbb Z$ time steps, a state vector
$\psi\in\mathcal H_A$ evolves according to
\begin{equation}
    \psi \longmapsto V^n\psi ,
\end{equation}
while an observable $A\in\mathcal A$ transforms in the Heisenberg picture as
\begin{equation}
    A\longmapsto V^{n}AV^{-n} .
\end{equation}

At this stage, the atom and the $R$-fields are independent systems. We also denote $\mathcal{E}^{(0)}_{\geq \tau}$ as  the algebra describing possible events and measurements of physical quantities at times $\geq \tau$. The superscript $0$ reminds us that we are in the situation without interaction between the `` atom'' and the $R$ fields. In particular, we have
\begin{align}
    \mathcal{E}^{(0)}_{\geq \tau}= \overline{\bigvee_{\tau' \geq \tau} \mathcal{A} \otimes \bigg(\bigotimes_{\tau' < \tau } \mathbb{1}_{\tau'} \otimes\mathcal{B}_{[\tau, \tau']} \otimes \bigotimes_{\tau' > \tau} \mathbb{1}_{\tau'} \bigg)}^w \simeq \mathcal{A} \otimes \mathcal{B}_{\geq \tau}
\end{align}
It is also obvious that PDP still holds
\begin{align}
    \mathcal{E}^{(0)}_{\geq \tau'} \subsetneq \mathcal{E}^{(0)}_{\geq \tau} \quad \textit{for} \: \: \tau' > \tau
\end{align}
To specify the time evolution, we introduce an embedding $*$ homomorphism $\alpha^{(0)}: \mathcal{E}^{(0)}_{\geq \tau +1} \rightarrow \mathcal{E}^{(0)}_{\geq \tau}$ by
\begin{align}
    \alpha^{(0)} (A \otimes E)=VA V^{*} \otimes \mathbb{1}_{\tau} \otimes E, \quad A \in \mathcal{A},  \: E \in \mathcal{B}_{\geq \tau +1}
\end{align}
Note that $VAV^{*} \in \mathcal{A}$ and $\mathbb{1}_{\tau} \otimes E \in \mathcal{B}_{\tau +1}$. Now, if $\omega_{\tau}$ is a state on the algbera $\mathcal{E}^{(0)}_{\geq \tau}$, then $\omega_{\tau +1}$ is simply given by $\omega_{\tau}\left.\right|_{\mathcal{E}^{(0)}_{\tau+1}}$: let $A \otimes E \in \mathcal{E}^{(0)}_{\geq \tau+1}$
\begin{align}
    \omega_{\tau +1} (A \otimes E) =  \omega_{\tau} (\alpha^{(0)}(A \otimes E))=  \omega_{\tau} (VAV^{*} \otimes \mathbb{1}_{\tau} \otimes E)
\end{align}

In the next subsection, we will introduce an interaction through which the
$R$-fields act on the atom and modify the evolution.

\subsection{Interaction,  Induced Dynamics and Decoherence }
We now study the interaction between the ``atom'' and the $R$-fields.
The coupled system is still described by a descending filtration of von Neumann
algebras $\{\mathcal{E}_{\geq \tau}\}_{\tau\in\mathbb{Z}}$
satisfying
\begin{equation}
    \mathcal{E}_{\geq \tau}
    \simeq
    \mathcal{A}\otimes \mathcal{B}_{\geq \tau}.
    \tag{11.30}
\end{equation}

To construct the interaction, we first introduce a family 
of mutually orthogonal projections  $\{Q_n\}_{n=0}^{K-1},
\:
K\leq M$ on $\mathcal{H}_A$, satisfying
\begin{equation}
    Q_n^*=Q_n,
    \qquad
    Q_nQ_m=\delta_{mn}Q_n,
    \qquad
    \sum_{n=0}^{K-1}Q_n
    =
    \mathbf{1}_{\mathcal{H}_A}.
    \tag{11.31}
\end{equation}
Thus, the family $\{Q_n\}_{n=0}^{K-1}$ defines a resolution of the identity
on the ``atomic'' Hilbert space. We also choose a family of $K$ unitary operators
$\{T^{(n)}\}_{n=0}^{K-1}
\subset
\mathcal{U}(\mathcal{H}_{\tau=0})$. For each $\tau$, we define the corresponding family $\{T_{\tau}^{(n)}\}_{n=0}^{K-1}
\subset
\mathcal{U}(\mathcal{H}_{\tau})$
by transporting the operators $T^{(n)}$ along the shift $S$:
\begin{equation}
    T_{\tau}^{(n)}
    =
    S T^{(n)} S^*.
    \tag{11.32}
\end{equation}
Now, we define a unitary operator $W_0$ acting on $\mathcal{H}_A \otimes \mathcal{H}_{\tau=0}$ as \footnote{One can easily check that $W_0$ is unitary 
\begin{align*}
   W_{0}^{*}W_{0}= W_{0}W_{0}^{*}= \sum_{n,m=0}^{K-1} Q_n Q_m \otimes T^{(n)}T^{(m)*}= \sum_{n,m=0}^{K-1} \delta_{nm}Q_m\otimes T^{(n)}T^{(m)*}=1 
\end{align*}}
\begin{align}
    W_0 = \sum_{n=0}^{K-1} Q_n \otimes T^{(n)}
\end{align}
Similarly, at each time $\tau$, we can also construct the unitary operator $W_{\tau}$ acting on $\mathcal{H}_A\otimes \mathcal{H}_{\tau}$ by
\begin{align}
    W_{\tau}= (\mathbb{1}\otimes S)^{\tau}W_0 (\mathbb{1} \otimes S)^{- \tau}= \sum_{n=0}^{K-1} Q_n \otimes T^{(n)}_{\tau}
\end{align}
We specify the interaction between the `` atom'' and the $R$-fields by specifying an embedding $*$ homomorphism $\alpha: \mathcal{E}_{\geq \tau +1} \rightarrow \mathcal{E}_{\geq \tau}$ as follows: let $C \in \mathcal{E}_{\geq \tau+1}$, then
\begin{align}
    \alpha (C)= W_{\tau} (V \otimes \mathbb{1})C (V^* \otimes \mathbb{1} )W^{*}_{\tau} \in \mathcal{E}_{\geq \tau}
\end{align}
More concretely, let $C = A \otimes E$ with $A \in \mathcal{A}$ and $E \in \mathcal{B}_{\geq \tau+1}$, then we have
\begin{align}
    \alpha (C)= \sum_{n,m=0}^{K-1} Q_n A Q^{*}_m \otimes T^{(n)}_{\tau} T^{(m)*}_{\tau} \otimes E
\end{align}
We note that $Q_n A Q_{m}^{*} \in \mathcal{A}$ and $T^{(n)}_{\tau} T^{(m)*}_{\tau} \otimes E \in \mathcal{B}_{\geq \tau}$ for each pair $n,m$. Now, if $\omega_{\tau}$ is a state on $\mathcal{E}_{\geq \tau}$, then $\omega_{\tau+1}= \omega_{\tau} \vert_{\mathcal{E}_{\geq \tau +1}}$ is given by
\begin{align}
    \omega_{\tau+1} (C)= \omega_{\tau} (\alpha (C)), \quad C \in \mathcal{E}_{\geq \tau +1}
\end{align}
In particular, we can choose the state $\omega_{\tau}$ on $\mathcal{E}_{\geq \tau} \simeq \mathcal{A} \otimes \mathcal{B}_{\geq \tau}$ to be 
\begin{align}
    \omega_{\tau} (A \otimes B)= \Tr (\Omega A) \cdot  \phi_{\Vec{k}} (B) = \Tr (\Omega A) \langle \Phi_{\Vec{k}} \vert B \vert \Phi_{\vec{k}}\rangle, \quad A \in \mathcal{A}, \: B \in \mathcal{B}_{\geq \tau}
\end{align}
where $\Omega$ is a density matrix on $\mathcal{H}_A$. So, for $C=A \otimes E \in \mathcal{A} \otimes \mathcal{B}_{\geq \tau+1} \simeq \mathcal{E}_{\geq \tau+1}$, it follows that
\begin{align}
    \omega_{\tau+1} (C)= \sum_{n,m=0}^{K-1} \Tr (V^{*}Q_m \Omega Q_n V  A) \cdot \langle \varphi_{k (\tau)} \vert T^{(n)}_{\tau} T^{(m)*}_{\tau} \ket{\varphi_{k(\tau)}} \cdot  \phi_{\vec{k}} (E) \label{important t+1 state}
\end{align}
We note the the second factor in (\ref{important t+1 state}) can be further written as
\begin{align}
\langle \varphi_{k (\tau)} \vert T^{(n)}_{\tau} T^{(m)*}_{\tau} \ket{\varphi_{k(\tau)}} = \sum_{a=0}^{N-1} \langle \varphi_{k (\tau)} \vert T^{(n)}_{\tau} \vert \varphi_{a} \rangle \langle \varphi_a \vert T^{(m)*}_{\tau} \vert \varphi_{k(\tau)} \rangle := \sum_{a=0}^{N-1} M^{(n)}_{a, k(\tau)}M^{(m)*}_{a, k(\tau)}
\end{align}
where $M^{(n)}_{a, k(\tau)}  =\langle \varphi_{k (\tau)} \vert T^{(n)}_{\tau} \vert \varphi_{a} \rangle $. We can also absorb the second factor in $(\ref{important t+1 state})$ into the first factor to obtain
\begin{align}
   \omega_{\tau+1} (C) &= \sum_{n,m=0}^{K-1} \sum_{a=0}^{N-1}\Tr  \bigg(V^* (M^{(m)}_{a, k(\tau)} Q_m)^* \Omega(M^{(n)}_{a, k(\tau)} Q_n) V   A
  \bigg)  \cdot \phi_{\Vec{k}} (E) \\
  &= \sum_{a=0}^{N-1} \Tr \bigg(V^* K^{(a)*}_{k(\tau)} \Omega K^{(a)}_{k(\tau) }V A \bigg) \cdot  \phi_{\vec{k}}(E) \\
  &= \Tr \bigg(  \Psi_{\tau}(\Omega) A \bigg) \cdot \phi_{\vec{k}} (E) \label{key discovery for discrete FP}
\end{align}
where we have defined 
\begin{align}
    K^{(a)}_{k(\tau)} = \sum_{n=0}^{N-1} M^{(n)}_{a, k(\tau)} Q_n, \quad \Psi_{\tau} (\Omega)= \sum_{a=0}^{N-1} V^{*}K^{(a)*}_{k(\tau)}\Omega K^{a}_{k(\tau)}V  \label{cptppp}
\end{align}
We note that the operators $K^{(a)}_{k(\tau)}$  and the map $\Psi_{\tau}$ satisfy  
\begin{align}
    \sum_{a=1}^{N-1}K^{(a)}_{k(\tau)} K^{(a )*}_{k(\tau)}=\sum_{a=0}^{N-1} M^{(n)}_{a, k(\tau)} M^{(n)*}_{a, k(\tau)}=1, \quad \Tr \Psi_{\tau} (\Omega)= \Tr (\Omega) 
\end{align}
Hence, by using that $V$ is unitary on $\mathcal{H}_A$, we know the map $\Psi_{\tau}$ is in the Kraus form and therefore it is a Completely Positive Trace Preserving (CPTP) map \footnote{A map $\Phi$ is called completely positive iff $\Phi \otimes\mathbb{1}_c$ is positive for all $c \in \mathbb{N}$. }.

\par Alternatively, we can infer the effective dynamics of the "atomic" part directly from (\ref{important t+1 state}). We define
\begin{align}
    g^{nm}_{\tau}= \langle \varphi_{k(\tau)} \vert T^{(n)}_{\tau} T^{(m)*}_{\tau} \vert \varphi_{k(\tau)} \rangle
\end{align}
So, when we march from $\tau$ to $\tau+1$, the effective evolution of the `` atom'' density matrix is
\begin{align}
    \Omega \rightarrow \sum_{n,m=0}^{K-1} g^{nm}_{\tau} V^{*} Q_m \Omega Q_n V   \label{effective dynamics}
\end{align}
In the \textbf{weak coupling limit}, i.e.,
\begin{align}
    g^{nm}_{\tau}= 1 + \mathcal{O} (\epsilon), \quad \epsilon \ll 1  \label{weak coupling limit}
\end{align} 
Substituting (\ref{weak coupling limit}) into (\ref{effective dynamics}), we obtain
\begin{align}
    \Omega_{\tau +1} &= \sum_{n,m=0}^{K-1}   V^{*}Q_n \Omega Q_m V + \mathcal{O}(\epsilon)\\
    &= \sum_{n,m=0}^{K-1} V^{*} \Omega V + \mathcal{O} (\epsilon)
\end{align}
So, we see that we recover Schr\"odinger evolution up to order $\mathcal{O}(\epsilon)$ corrections. Put it differently, the weak coupling limit condition (\ref{weak coupling limit}) implies that $T^{(m) *}_{\tau} \ket{\varphi_{k (\tau)}} =T^{(n)*}_{\tau} \ket{\varphi_{k(\tau)}}+ \mathcal{O}(\epsilon), \forall m,n =0, \cdots, K-1$.
This means that the states of the environmental degrees of freedom in $\mathcal{H}_{\tau}$ that become entangled  with the `` atom'' due to the interaction are nearly indistinguishable for different $ Q_n$ sectors, resulting in negligible decoherence.

It is also instructive at this stage to impose sharper condition, namely that we let the duration for each time step $dt= \epsilon$ and we let the continuous time $t= \epsilon \tau$ where $\tau$ here is the discrete time. In such a setup, we can write 
\begin{align}
    \Omega_{t + \epsilon} = \sum_{n,m=0}^{K-1} g^{nm}(\epsilon) V^{*}(\epsilon)Q_m \Omega_t Q_n V(\epsilon), \quad V(\epsilon)= e^{i H_A \epsilon} \label{intsructive cont lim}
\end{align}
where $ H_A $ is the canonical Hamiltonian on $\mathcal{H}_A$. By definition $\{g^{nm} (\epsilon)\}$ is a Gram matrix and thus it must be positive semidefinite \footnote{Note that $g^{nm}(\epsilon)= \langle T_{\frac{t}{\epsilon}}^{(n)*}(\epsilon) \varphi  \vert T^{(n)*}_{\frac{t}{\epsilon}}(\epsilon) \varphi \rangle$. It satisfies $g^{nn} (\epsilon)=1, \overline{g^{nm} (\epsilon)}= g^{mn}(\epsilon)$ and $g(\epsilon) \geq 0$. }. Moreover, as the time step $\epsilon \rightarrow 0$, we have $  g^{nm}(\epsilon) \rightarrow 1$. So, we can express $g^{nm} (\epsilon)$ in the weak coupling limit as
\begin{align}
    g^{nn}(\epsilon)=1, \quad g^{nm} (\epsilon)= 1- \epsilon \Gamma^{nm} + \mathcal{O}(\epsilon^2) \: \textit{for} \: \; n \neq m
\end{align}
where $ \overline{\Gamma^{mn }}= \Gamma^{nm}$. For simplicity, we can take all $\Gamma^{nm}= \gamma, m \neq n$. In particular, we should interpret $\gamma$ here as decoherence rate.
With such considerations, (\ref{intsructive cont lim}) becomes \footnote{In deriving GKSL equation, we use $\sum_{n \neq m} Q_n \Omega_t Q_m=\Omega_t - \sum_{n}Q_n \Omega_t Q_n= \sum_n \frac{1}{2}\{  Q_n, \Omega_t\}- \sum_n Q_n \Omega_t Q_n$.}
\begin{align}
    \Omega_{t + \epsilon}= \Omega_t + \epsilon \bigg(-i [H_A, \Omega_t] + \gamma \sum_{n=0}^{K-1} \big( Q_n \Omega_t Q_n- \frac{1}{2} \{ Q_n, \Omega_t \}    \big) \bigg) + \mathcal{O} (\epsilon^2)
\end{align}
Equivalently, we obtain
\begin{align}
    \frac{\partial \Omega_{t}}{\partial t}= \mathcal{L}_{\gamma} (\Omega_t), \quad \mathcal{L}_{\gamma} (\Omega_t)=-i [H_A, \Omega_t] + \gamma \sum_{n=0}^{K-1}\big ( Q_n \Omega_t Q_n- \frac{1}{2} \{ Q_n, \Omega_t \}  \big)  \label{Q Lindblad equation}
\end{align}
which is just the standard Gorini-Kossakowski-Sudarshan-Lindblad (GKSL) equation in the study of open quantum system with Lindblad operators given by $L_n = Q_n$. Hence, we see that in the weak coupling limit, the effective dynamics of the `` atomic'' part in the \textbf{continuous time limit }is governed by  GKSL equation.

\par In contrast, in the \textbf{strong coupling limit},
\begin{align}
    g^{nm}_{\tau} = \delta^{nm} + \mathcal{O} (\epsilon), \quad \epsilon\ll 1
\end{align}
we obtain
\begin{align}
    \Omega_{\tau +1} = V^{*} \big( \sum_{n=0}^{K-1} Q_n \Omega_\tau Q_n  \big ) V+ \mathcal{O}(\epsilon)
\end{align}
In particular, the off-diagonal coherences are suppressed and only the block diagonal part of the density matrix is retained. Moreover, the strong-coupling limit can also be understood as the large-dephasing-rate limit of GKSL equation. We now project the GKSL equation (\ref{Q Lindblad equation}) we derived before into one particular $n \neq m$ sector
\begin{align}
    \frac{\partial ( Q_n\Omega_tQ_m)}{\partial t}= -i Q_n[H_A, \Omega_t]Q_m - \gamma  Q_n \Omega_t Q_m, \quad n \neq m
\end{align}
In the strong coupling limit, the decoherence rate $\gamma$ is much larger than the characteristic Hamiltonian frequency \footnote{The  characteristic Hamiltonian frequency is the typical rate at which the Hamiltonian changes the state through unitary evolution.}, i.e., $\gamma \gg \vert \vert H_A \vert \vert _{\mathcal{H}_A} $.  So, the decoherence time scale $\tau_{dec}$ \footnote{The decoherence timescale is the characteristic time over which off-diagonal coherences become substantially suppressed.} is much shorter than the Hamiltonian timescale $\tau_{H_A} $\footnote{The Hamiltonian timescale is the characteristic time over which the unitary evolution generated by \(H\) changes the state appreciably.}
\begin{align}
    \tau_{dec} \sim \frac{1}{\gamma} \ll \frac{1}{\vert \vert H_A \vert \vert_{\mathcal{H}_A}} \sim \tau_{H_A}
\end{align}
Hence, during a very short time interval $t \sim \tau_{dec}$, we can safely neglect the contribution from the Hamiltonian term and write $\frac{\partial (Q_n \Omega_t Q_m)}{\partial t} \approx - \gamma Q_n \Omega_t Q_m$ which leads to 
\begin{align}
    Q_n \Omega_t Q_m \approx e^{- \gamma t} Q_n \Omega_0 Q_m, \quad n \neq m, \: t \sim \tau_{dec}
\end{align}
Therefore, after an infinitesimal time step $\epsilon$, we also have $Q_n \Omega_{\epsilon} Q_m \sim e^{- \gamma \epsilon}Q_n \Omega_0 Q_m, \: n \neq m$. This means that the corresponding Gram matrix is
\begin{align}
    g^{nn} (\epsilon)=1, \quad g^{nm}(\epsilon)=e^{- \gamma \epsilon}, \; n \neq m
\end{align}
For the weak coupling limit, we have $ \gamma\epsilon \ll 1 $ so we have $e^{- \gamma \epsilon} = 1 - \gamma \epsilon+ \mathcal{O}(\epsilon^2) $. However, in the strong coupling limit, we have $\gamma \epsilon \gg 1$ so we have $ g^{mn} (\epsilon) \approx 0, \: n \neq m$ which leads to that $g^{nm}(\epsilon) \approx \delta^{nm}$. So, we see that the strong coupling limit can be viewed as $\gamma \epsilon \rightarrow \infty$ limit generated by the dephasing Lindbladian.

To conclude this subsection, we want to emphasize the key role of the
Principle of Diminishing Potentialities (PDP) in governing the dynamics of
this model. We see that, as we evolve from time $\tau$ to $\tau+1$, turning
on the interaction generates entanglement between the ``atomic'' degree of
freedom and the degrees of freedom in $\mathcal H_\tau$ of the $R$-field
system in the global state. For instance, if the global state at time
$\tau$ is given by
\begin{equation}
    \omega_\tau(A\otimes E)
    =
    \Tr(\Omega A)\cdot \phi_{\vec{k}}(E),
    \qquad
    A\in\mathcal A,\quad E\in\mathcal B_{\geq\tau},
\end{equation}
then, due to the interaction, the global state at time $\tau+1$ becomes
\begin{equation}
    \omega_\tau\bigl(\delta(A\otimes E)\bigr)
    =
    \Tr\!\bigl(\Psi_\tau(\Omega)A\bigr)\cdot
    \phi_{\vec{k}}(S E S^{*}) = \Tr\!\bigl(\Psi_\tau(\Omega)A\bigr)\cdot
    \phi_{\sigma{(\vec{k})}} (E)
\end{equation}
Here $\delta$ is a $*$-endomorphism, i.e., $\delta:
    \mathcal A\otimes\mathcal B_{\geq\tau}
    \longrightarrow \mathcal{E}_{\geq \tau+1} \simeq
    \mathcal A\otimes\mathcal B_{\geq\tau+1}$ defined by $\omega_{\tau } \circ \delta (\cdot_{\mathcal{A}} \otimes \cdot_{\mathcal{B}_{\geq \tau}})= \Tr \big( \Psi_{\tau} (\Omega) \cdot_{\mathcal{A}}  \big) \cdot \phi_{ \sigma(\vec{k})} ( \cdot_{\mathcal{B}_{\geq \tau}})$. 
Due to the interaction, the ``atomic'' system  and the $R$-field subsystem
associated to  $\mathcal H_\tau$  in the global state
$\omega\circ\delta$ are mixed together.  However, at time $\tau+1$, the algebra accessible to us
is only $\mathcal E_{\geq\tau+1}
    \simeq
    \mathcal A\otimes\mathcal B_{\geq\tau+1}$. Since $\mathcal B_{\geq\tau+1}
    \subsetneq
    \mathcal B_{\geq\tau}$, it also follows that $\mathcal E_{\geq\tau+1}
    \subsetneq
    \mathcal E_{\geq\tau}$.
Therefore, the restricted state $\omega_\tau|_{\mathcal E_{\geq\tau+1}}$
becomes  mixed because we can no longer access the information
carried by the degrees of freedom in $\mathcal H_\tau$ and then the part of the state $\omega \circ \delta$ belonging to $\mathcal{H}_{\tau}$ has to be traced out and absorbed into the `` atomic''system. This loss of access arises precisely
from the shrinking of the algebras $\mathcal E_{\geq\tau+1}
    \subsetneq
    \mathcal E_{\geq\tau}$.

\subsection{The Ontology of the Evolution} \label{sec: ontology}
In this subsection, we want to discuss the ontology of the dynamics of FP model which is based on the mechanism for the occurrence of events in the ETH approach to quantum mechanics \cite{Frohlich:2026epd} \cite{Frohlich:2024kyo}. We recall that whether an event occurs or not depends on the center of the centralizer of the state. We  discuss the `` atomic part'' first. Suppose at time $\tau$ the state  of the `` atomic system'' is described by a density matrix $\Omega_{\tau}$, then its centralizer $\mathcal{C}_{\Omega_\tau}$ is given by

\begin{align}
    \mathcal{C}_{\Omega_\tau}= \{ A \in \mathcal{A}  \vert  [A, \Omega_{\tau}]=0 \}
\end{align}
Its center $Z_{\Omega_\tau}= \mathcal{C}_{\Omega_\tau} \cap \mathcal{C}_{\Omega_\tau}'$ is  given by the bounded function of the density matrix $\Omega_\tau$
\begin{align}
    Z_{\Omega_\tau}= \{f(\Omega_\tau)  \vert f \: \textit{is bounded} \}
\end{align}

For the $R$-field sector, we choose the stationary product state to be $\omega_{\beta \rightarrow \infty}= \phi_{\vec{0}}$ which is a pure product state. Its GNS representation is irreducible, and the corresponding von Neumann algebra $\mathcal{B}_{\geq \tau} \simeq \mathcal{B}(\mathcal{H}_{\infty})$ is a Type $\RN{1}_{\infty}$ factor. We recall that $\ket{\Phi_{\vec{0}}}$ is the GNS vector representing the zero-temperature product state. We define the rank-one projection as $P_{\vec{0}} = \ket{\Phi_{\vec{0}}} \bra{\Phi_{\vec{0}}}$. For a Type $\RN{1}$ factor $\mathcal{B}_{\geq \tau}$, the centralizer of such a vector state is the commutant of its density matrix $P_{\vec{0}}$
\begin{align}
    \mathcal{C}_{\vec{0}} (\mathcal{B}_{\geq \tau})= \{A \in \mathcal{B}(\mathcal{H}_{\infty}) \vert [A, P_{\vec{0}}]=0  \}
\end{align}
Equivalently, with respect to the decomposition $\mathcal{H}_{\infty}= \mathbb{C} \ket{\Phi_{\vec{0}}} \oplus \mathcal{H}^{\perp}_{\infty}$ where $\mathcal{H}^{\perp}_{\infty}=  P_{\vec{0}}^{\perp}\mathcal{H}_{\infty}$ and $P^{\perp}_{\vec{0}}= 1- P_{\vec{0}}$, the centralizer takes the block-diagonal form
\begin{align}
    \mathcal{C}_{\vec{0}} (\mathcal{B}_{\geq \tau}) \simeq \mathbb{C} P_{\vec{0}} \oplus \mathcal{B}(\mathcal{H}^{\perp}_{\infty})
\end{align}
Its center is therefore
\begin{align}
    Z_{\vec{0}} (\mathcal{B}_{\geq \tau})= \mathbb{C} P_{\vec{0}} \oplus \mathbb{C} (\mathbb{1}-P_{\vec{0}})
\end{align}
Now, for the combined system, the relevant algebra is $\mathcal{E}_{\geq \tau+1} \simeq \mathcal{A} \otimes  \mathcal{B}_{\geq \tau +1} $ given by
\begin{align}
    \mathcal{E}_{\geq \tau +1}= \sum_{m,n} VQ_m \mathcal{A} Q_nV^* \otimes T^{(m)}_{\tau} T^{(n)*}_{\tau} \otimes \mathcal{B}_{\geq \tau +1}  \label{explicit E algebrasss}
\end{align}
The state $\omega_{\tau +1}$ for this algebra $\mathcal{E}_{\geq \tau +1}$ is given by $\omega_{\tau}\vert _{\mathcal{E}_{\geq \tau +1}}= \Tr (\Omega_{\tau} \cdot_{\mathcal{A}}) \otimes \phi_{\vec{0}} $. From (\ref{key discovery for discrete FP}), we know the pair $(\mathcal{E}_{\geq \tau +1}, \omega_{\tau }\vert_{\mathcal{E}_{\geq \tau +1}})$  is completely physically equivalent to the  pair  $\bigg[\mathcal{A} \otimes \mathcal{B}_{\geq \tau +1}, \Tr (\Psi_{\tau}(\Omega_\tau) \cdot_{\mathcal{A}})  \phi_{\vec{0}}(\cdot_{\mathcal{B}_{\geq \tau +1}}) \bigg] $ \footnote{
The pair $\left(
\mathcal{E}_{\geq \tau+1},
\operatorname{Tr}(\Omega_\tau\,\cdot_{\mathcal A})
\otimes \phi_{\vec{0}}
\right)$
is expressed in the Heisenberg picture, whereas
$\left(
\mathcal A\otimes\mathcal B_{\geq\tau+1},
\operatorname{Tr}\bigl(\Psi_\tau(\Omega_\tau)\,
\cdot_{\mathcal A}\bigr)
\otimes\phi_{\vec{0}}
\right)$
is its corresponding Schr\"odinger-picture description.
}. So, to capture the ontology of the dynamics when we go from $\tau$ to $\tau +1$, we can just work with the product algebra $\mathcal{A} \otimes \mathcal{B}_{\geq \tau +1}$ equipped with the state  $ \Psi_{\tau } (\Omega_\tau) \otimes P_{\vec{0}}$.  We know that the centralizer of the state $\Psi_{\tau} (\Omega) \otimes P_{\vec{0}}$ on the algebra $\mathcal{A} \otimes \mathcal{B}_{\geq \tau +1}$ is the collection of operators that commute with $\Psi_{\tau } (\Omega_\tau) \otimes P_{\vec{0}}$. From our discussion before, we know it is
\begin{align}
    \mathcal{C}_{\tau +1} (\mathcal{A} \otimes \mathcal{B}_{\geq \tau +1})= \{ \mathcal{C}_{\Psi_{\tau}(\Omega_{\tau})} \otimes P_{\vec{0}} \oplus \mathcal{B} (\mathcal{H}_{A} \otimes \mathcal{H}^{\perp}_{\infty} )  \} 
\end{align}
Its center is
\begin{align}
    Z_{\tau +1} (\mathcal{A} \otimes \mathcal{B}_{\geq \tau +1})= \{ f(\Psi_{\tau}(\Omega_{\tau})) \otimes  P_{\vec{0}} \oplus \mathbb{1}_{\mathcal{A}} \otimes (1- P_{\vec{0}}) \vert \: f \: \textit{bounded} \}
\end{align}

The second term is irrelevant at all in the state under consideration because
\begin{align}
    \phi_{\vec{0}} ((1-P_{\vec{0}})B)=0, \quad \forall B \in \mathcal{B}_{\geq \tau}
\end{align}
Hence, at time $\tau +1$, the cental projections carrying a nonzero probability are entirely determined by the spectral structure of the atomic density matrix $\Psi_\tau (\Omega_\tau)$. So, according to the Collapse Postulate (CP) of the ETH approach, one of the spectral projectors of $\Psi_{\tau} (\Omega_\tau)$ will be selected as the actual event at time $\tau +1$ and the collapse of state occurs. This process is repeated for all time steps. In this sense, the evolution of the events structure  reduces entirely to the evolution of the `` atomic'' part \footnote{In the original physical motivation
of the FP model, the nonrelativistic limit
$c\to\infty$ implies that photons emitted at
earlier times have already escaped to spatial
infinity. Consequently, the degrees of freedom
associated with the previously emitted radiation
no longer belong to the algebra of future
potentialities. If the incoming $R$-fields
are initially prepared in their vacuum-like
reference state, the future environmental
degrees of freedom remain unaffected by
the previous interactions.
}.

\par To make our discussion more precise, we first start with the effective one-step dynamics,
\begin{align}
\Omega_{\tau}\otimes P_{\vec 0}
\longmapsto
\Psi_{\tau}(\Omega_{\tau})\otimes P_{\vec 0}, \quad \Omega_{\tau +1}= \Psi_{\tau} (\Omega_{\tau})
\end{align}
 In the continuous-time limit, writing $t=\tau\,dt$, the reduced evolution is generated by the GKSL equation,
\begin{align}
\Omega_{t+dt}
=
\Omega_t
+
dt\,\mathcal{L}_{\gamma}(\Omega_t)
+
\mathcal{O}(dt^2),   \label{GKSL222}
\end{align}
where $\mathcal{L}_{\gamma}$ denotes the corresponding Lindblad generator, with decoherence rate $\gamma$ in the model considered here. Since $\Omega_{t+dt}$ is a positive self-adjoint operator, it admits a spectral decomposition by diagonalizing (\ref{GKSL222})
\begin{align}
\Omega_{t+dt}
=
\sum_{\alpha=0}^{M-1}
p_{\alpha}(t+dt)\,
P_{\alpha}(t+dt),
\end{align}
where the $p_{\alpha}$ are the distinct eigenvalues and $P_{\alpha}$ the associated spectral projections. It follows that the physically relevant center of the centralizer is generated by the family of projectors $\{P_{\alpha}(t+dt)\otimes P_{\bar 0}\}_{\alpha=0}^{M-1}$. These projections define the family of possible event alternatives at time $t+dt$. An actual event corresponds to the realization of one member of this family, according to the event-selection rule of the ETH framework. To further understand this, we consider the situation that the state $\Omega_t= \Pi_t$ which is a pure state. At time $t + dt$, we have
\begin{align}
    \Omega_{t+ dt}= \Pi_t + dt \: \mathcal{L}_{\gamma} (\Pi_{t}) + \mathcal{O} (dt^2)  \label{GKSL 3333}
\end{align}
In general, after diagonalizing  (\ref{GKSL 3333}), the universal pattern is
\begin{align}
    \Omega_{t+dt} =p_{nj}[t, t+dt] \hat{\Pi}_0  + \sum_{\delta=1}^{M-1} p_{\delta}[t, t+ dt] \hat{\Pi}_{\delta}
\end{align}
where $p_{nj}[t, t+dt] \approx 1 - \mathcal{O}(dt) $ is the probability with no jump and $\hat{\Pi}_0 \approx \Pi_t$. So, the difference between $\hat{\Pi}_0$ and $\Pi_t$ is almost negligible.  Furthermore, $p_\delta[t, t+dt] \approx \mathcal{O}(dt)$ denotes the probability with a jump which is very small.  The state $\hat{\Pi}_{\delta}$ is significantly different from $\hat{\Pi}_0$. To be able to understand why this pattern is general, a useful trick is to use first-order perturbation theory. Since $\Pi_t$ is a pure state, its spectrum $spec(\Pi_t)=(1, 0 , \cdots, 0)$ and we can view it as a free Hamiltonian $H_{free}$. We treat the time step $dt$ as an infinitesimal parameter and we take the Lindbladian $\mathcal{L}_{\gamma} (\Omega_t)$ as a perturbation. We can apply the machinery of the perturbation theory to obtain the leading order correction of the leading eigenvalue of $\Pi_t$ and also the leading order correction to its eigenvector. A straightforward manipulations give \footnote{Let $H(\epsilon)= H_{free} + \epsilon V$. Suppose that $H_{free}$ has a simple eigenvalue $E_0$ with normalized eigenvalue $\ket{\psi_0}$. We define $\Pi = \ket{\psi_0}\bra{\psi_0}, \Pi^{\perp}= 1- \Pi$ and we have $\Pi \ket{\psi_0}= \ket{\psi_0}$. With perturbation $V$ included, the updated leading eigenvalue is $\ket{\psi (\epsilon)}= \ket{\psi_0} + \epsilon \ket{\psi_1} + \mathcal{O} (\epsilon^2) $
. Recall that $\psi_1$ is given by the formula
\begin{align*}
    \ket{\psi_1}= \sum_{m \neq 0} \frac{\bra{m} V \ket{\psi_0}}{E_0 - E_m}\ket{m}= (E_0- H_{free})^{-1} \Pi^{\perp} V \ket{\psi_0}
\end{align*}
So, its corresponding density matrix is 
\begin{align*}
    \Pi (\epsilon)= \ket{\psi (\epsilon)} \bra{\psi (\epsilon)}= \Pi + \epsilon \bigg( \Pi V \Pi^{\perp} (E_0 - H_{free})^{-1} + (E_0 - H_{free})^{-1} \Pi^{\perp} V \Pi  \bigg) + \mathcal{O} (\epsilon^2)
\end{align*}
Note that the term in the big bracket is $\ket{\psi_0} \bra{\psi_1} + \ket{\psi_1} \bra{\psi_0}$ and it can be written as $[\Pi, S]$ where $S$ is an antisymmetric operator given by $S=- (E_{0} - H_{free})^{-1} \Pi^{\perp} V \Pi + \Pi V \Pi^{\perp} (E_{0} - H_{free})^{-1}= - \ket{\psi_1} \bra{\psi_0} + \ket{\psi_0} \bra{ \psi_1}$. So, it follows that $[\Pi, S]= \ket{\psi_0} \bra{\psi_1}+ \ket{\psi_1} \bra{\psi_0}$. Moreover, the standard trick also gives $E(\epsilon)= E_0 +  \epsilon \bra{\psi_0} V \ket{\psi_0} + \mathcal{O}(\epsilon^2)$ So, we see that we obtain
\begin{align*}
    \Pi(\epsilon)= \Pi - \epsilon[S, \Pi] +\mathcal{O}(\epsilon^2), \quad E (\epsilon)= E_0 + \epsilon \:\Tr (\Pi V) + \mathcal{O}(\epsilon^2)
\end{align*}}
\begin{align}
    p_{nj}[t, t+ dt]= 1 + dt \; \Tr[\Pi_t \mathcal{L}_{\gamma}(\Pi_t)] \label{no jump probability time step}
\end{align}
One can verify that $\Tr[\Pi_t \mathcal{L}_{\gamma}(\Pi_t)]<0$ by using Cauchy-Schwarz inequality. This implies that the probability of no jump for a finite time interval $[t_0, t_1]$ is
\begin{align}
    p_{nj}[t_0, t_1]= \exp  \bigg( \int_{t_0}^{t_1} \Tr [\Pi_s \: \mathcal{L}_{\gamma}(\Pi_s)]  \;ds\bigg) <1 \label{no jump prob}
\end{align}
and the dynamical equation governing the density matrix $\Pi_{t}$  \textbf{in the absence of quantum jump} is
\begin{align}
    \frac{\partial \Pi_t}{\partial t}= \Pi_t \mathcal{L}_{\gamma}(\Pi_t) \Pi^{\perp}_t + \Pi_t^{\perp} \mathcal{L}_{\gamma}(\Pi_t) \Pi_t   \label{absence of jump eqn}
\end{align}
Moreover, we can also conjugate $\Omega_{t+ dt}$ with $\Pi^{\perp}_{t}$ to obtain
\begin{align}
    \Pi^{\perp}_t \Omega_{t+ dt} \Pi^{\perp}_{t}= dt \: \Pi^{\perp}_{t} \mathcal{L}_{\gamma} (\Pi_t) \Pi^{\perp}_{t}
\end{align}
Hence, quantum jump probabilities $p_{\delta}[t+ dt] \sim \mathcal{O}(dt), \: \delta= 1, \cdots, M-1$ can be obtained by diagonalizing the matrix $dt \:   \Pi^{\perp}_t \Omega_{t+ dt} \Pi^{\perp}_t$.

\par Now, we can also understand why in the previous subsection we see that for a very short duration of time, the effective dynamics of the `` atom'' is given by Schr\"odinger evolution up to small corrections in the \textbf{weak coupling limit}. After diagonalizing the density matrix $\Omega_{t + dt}$, we see that there is an overwhelming probability $p_{nj}[t, t+dt] \sim1 - \mathcal{O}(dt)$ for the Nature to pick $\hat{\Pi}_0 \sim \Pi_t$ as an actual event. This means that when we progess from $t$ to $t+dt$, there is an overwhelming probability that any change is negligible. Suppose this process occurs for $n$ time steps and when $n$ is sufficiently large, the accumulative probability $p_{nj}[t, t+ n \: dt] \sim (1- dt)^n \sim \exp (-n \: dt)$ becomes exponentially small. Therefore, for a significantly long duration of time, a continuous sequence of no quantum jump becomes unlikely and a quantum jump is likely to occur\footnote{The author learned this intuition from an online lecture by J\"urg Fr\"ohlich.}.

\section{DSW Horizon Decoherence}

\subsection{Causal Structure of Bifurcate Killing Horizon}

We begin by reviewing the causal structure of a spacetime
admitting a bifurcate Killing horizon. We consider a metric
of the form
\begin{align}
    g = -2 A(UV,x)\,dU\,dV
    + h_{ab}(UV,x)\,dx^a dx^b,
    \label{eq:bifurcate_metric}
\end{align}
where $U$ and $V$ are null coordinates, and $x^a$ denote
coordinates on the transverse spatial sections. We assume
that the coordinate chart extends across $U=0$ and $V=0$,
so that both coordinates may take positive and negative
values. The functions $A(UV,x)$ and $h_{ab}(UV,x)$ are
assumed to be smooth across $UV=0$, with $A(0,x)>0$
and $h_{ab}(0,x)$ to be positive definite.

\par We consider the vector field $\xi$ as
\begin{align}
    \xi= \kappa (V\partial_V - U \partial_U)
\end{align}
where $\kappa$ is a constant that will be identified
with the surface gravity. The one-parameter flow generated
by $\xi$ acts as $V \longmapsto e^{\kappa s}V,
    \:
    U \longmapsto e^{-\kappa s}U,
    \:
    x^a \longmapsto x^a$.
 Under this transformation, $UV$, $dU\,dV$, and $dx^a$
are invariant. Since the metric coefficients depend on
$U$ and $V$ only through their product $UV$, the metric
is invariant under the flow. Consequently, we have  $\mathcal{L}_{\xi}g=0$ and thus $\xi$ is a Killing vector field. Moreover, we have
\begin{align}
    \xi^2 = g_{\mu \nu} \xi^{\mu} \xi^{\nu}=2 \kappa^2 A UV
\end{align}
Thus, wherever $A>0$, the Killing vector field is
timelike in the regions $UV<0$, spacelike in the regions
$UV>0$, and null on the hypersurfaces $U=0$ and $V=0$. Now, we show that these two null hypersurfaces are Killing
horizons generated by $\xi$. We denote $ \Sigma_+ = \{U=0\}$ and $\Sigma_- = \{V=0\}$.
We recall that a vector field $X$ is tangent to a hypersurface
defined by $F=0$  when $X(F)\big|_{F=0}=0$.
For the Killing vector field $\xi$, we have $\xi(U)=-\kappa U$ and $\xi(V)=\kappa V$.
It follows  that $\xi(U)\big|_{\Sigma_+}=0$ and $\xi(V)\big|_{\Sigma_-}=0$.
Hence $\xi$ is tangent to both hypersurfaces. Moreover, the normal covector field for the null hypersurface $\Sigma_+$ and $\Sigma_-$ is $dU$ and $dV$ respectively. The corresponding normal vector fields are
\begin{align}
    n_{+}= - \frac{1}{A}\partial_V, \: n_-=- \frac{1}{A}\partial_U
\end{align}
Hence, we see that $n_{+} \parallel \xi \vert_{\Sigma_+}$ and $n_{-} \parallel \xi \vert_{\Sigma_-}$. Therefore, $\xi$ is both tangent and normal on the null hypersurfaces $\Sigma_+$ and $\Sigma_-$.
So, these hypersurfaces are Killing
horizons generated by $\xi$. The two Killing horizons intersect at the bifurcation
surface $\mathfrak{ B}=\{U=V=0\}$, where $\xi$ vanishes. It is also straightforward to check that
\begin{align}
    \nabla_{\xi} \xi = \pm \kappa \xi  , \quad \textit{on} \: \: \Sigma_{\pm}
\end{align}

\par Now, we will mainly focus on the wedge $U<0, V>0$. In fact, the one-parameter isometric flow generated by $\xi := \partial_t$ is given by $V(t)= V_0 e^{\kappa t}$ and $U(t)= U_0 e^{- \kappa t}$. We may choose the constants such that $-U_0= V_0=R >0$, so that 
\begin{align}
    t= \frac{1}{2 \kappa} \log \big( \frac{V}{-U} \big), \quad R=\sqrt{-UV}, \quad V= R e^{\kappa t}, \quad U= -R e^{- \kappa t}
\end{align}
It leads to that $dU dV=- dR^2 + \kappa^2 R^2 dt^2$. Therefore, we can write down the metric for the wedge $U<0, V>0$ as
\begin{align}
    g= 2A(-R^2, x) \bigg( -\kappa^2 R^2 dt^2 + dR^2   \bigg)+ h_{ab}(-R^2, x) dx^a dx^b
\end{align}
Note that the lapse function is $N= \sqrt{- \xi^2}= \kappa R \sqrt{2A}$. At fixed finite time $t$, $R \rightarrow 0$ approaches the bifurcation surface $\mathfrak{B}$. Approaching $U=0, V >0$ requires $t \rightarrow \infty$ and $R \rightarrow 0$ together. We also note that the Killing time $t$ is in the logarithmic relation with the affine null time $V$, i.e., $t \sim \log V$. Now, suppose that $A(0,x)= A_0$ i.e., when $R \rightarrow 0, A$ becomes independent of any coordinates\footnote{This condition is satisfied by a wide class of spacetimes admitting bifurcate horizons.}.  Around $R^2=0$, we have $A(-R^2,x) = A_0 + \mathcal{O}(R^2)$. So, in the near-horizon limit, we have
\begin{align}
    g \simeq - \kappa^2 \rho^2 dt^2 + d \rho^2 + h_{ab}(0, x) dx^a dx^b, \quad  \rho= \sqrt{2 A_0} R
\end{align}
By the standard Euclidean trick, we know that the temperature $T_K$ associated to the Killing time $t$ is $ T_K= \frac{\kappa}{2 \pi}$. The local Tolman temperature after taking into  account the gravitational blueshift effect is 
\begin{align}
    T_{\textit{local}}= \frac{\kappa}{2 \pi \sqrt{-2A (UV,x) UV}}
\end{align}
\subsection{Central Arguments}
In this subsection, we revisit the central arguments of Danielson, Satishchandran and Wald (DSW) regarding decoherence induced by Killing horizons \cite{Danielson:2022sga} \cite{Danielson:2022tdw}. We will mainly focus primarily on the Rindler horizon in Minkowski space, working out the relevant details explicitly. Other cases such as the event horizon in Schwarzschild spacetime and the cosmological horizon in de Sitter space proceed in a completely analogous manner. DSW consider a stationary
spacetime $(\mathcal{M},g_{\mu\nu})$ possessing a Killing
vector field $\xi^\mu$ which generates a Killing horizon
$\mathcal{H}$. We shall focus on the electromagnetic
case, although a similar analysis applies to the
gravitational field. They consider a charged  particle localized in the
stationary region where $\xi^\mu$ is timelike. This charged particle  is sent through a
Stern-Gerlach like apparatus, which prepares it in a
coherent superposition of two spatially separated wave
packets
 \begin{equation}
    |\Psi\rangle
    =
    \frac{1}{\sqrt{2}}
    \left(
        |\psi_1\rangle+|\psi_2\rangle
    \right).
\end{equation}
where for simplicity, $|\psi_1\rangle$ and
$|\psi_2\rangle$ are taken to be orthogonal,
unit-normalized wave packets. The two branches are maintained at distinct spatial
positions for a finite duration $T$, after which
they are recombined. During the intermediate stage,
the corresponding charge distributions are stationary
with respect to the Killing flow generated by
$\xi^\mu$. The duration $T$ is defined with respect
to the proper time of the stationary apparatus
maintaining the superposition. Each branch of the superposition gives rise to a
distinct electromagnetic current $j_i^\mu$,
$i=1,2$, and hence to a corresponding electromagnetic
field configuration $F_{\mu\nu}^{(i)}$ satisfying
Maxwell's equations $\nabla_\mu F^{(i)\mu\nu}
    =
    j_i^\nu, \: i=1,2$. The difference between the two field configurations i.e., $\Delta F_{\mu\nu}
    =
    F_{\mu\nu}^{(1)}
    -
    F_{\mu\nu}^{(2)}$
encodes information about the spatial separation
of the particle. Although the charge distributions are stationary
during the holding stage, their electromagnetic
fields extend to the Killing horizon. Consequently,
the two branches generally induce different
electromagnetic field configurations on the horizon. The interaction between the charged particle and
the quantized electromagnetic field leads to an
entangled state of the schematic form
\begin{equation}
    |\chi\rangle
    =
    \frac{1}{\sqrt{2}}
    \left(
        |\psi_1\rangle\otimes|\Phi_1\rangle
        +
        |\psi_2\rangle\otimes|\Phi_2\rangle
    \right),
\end{equation}
where $|\Phi_i\rangle$ denotes the electromagnetic
field state associated with the $i$-th branch.  After tracing over the electromagnetic field part, we obtain a density matrix for the charged particle 
\begin{align}
    \rho_A
    =
    \operatorname{Tr}_{\mathrm{EM}}
    \left(|\chi\rangle\langle\chi|\right)
    \nonumber
    =
    \frac{1}{2}
    \begin{pmatrix}
        1 & \langle\Phi_2|\Phi_1\rangle\\
        \langle\Phi_1|\Phi_2\rangle & 1
    \end{pmatrix}
\end{align}
where the matrix is expressed in the basis
$\{|1\rangle,|2\rangle\}$. The coherence between the two branches is therefore
characterized by the overlap of their corresponding
electromagnetic field states. To quantify the resulting decoherence, we briefly
review the method used by DSW. Let $\mathscr{S}$
denote the real vector space of smooth, source-free
solutions to Maxwell's equations, modulo gauge
transformations and subject to suitable boundary
conditions. For two solutions $A_a^{(1)},A_a^{(2)}
\in\mathscr{S}$, the symplectic form is defined by
\begin{equation}
    \Omega(A^{(1)},A^{(2)})
    =
    \int_{\Sigma} d\Sigma_a\,
    \left(
        A_b^{(1)}F^{(2)ab}
        -
        A_b^{(2)}F^{(1)ab}
    \right),
    \label{eq:EM_symplectic}
\end{equation}
where $\Sigma$ is a Cauchy surface and
$F_{ab}^{(i)}=2\nabla_{[a}A_{b]}^{(i)}$.
By virtue of Maxwell's equations, the symplectic
form is independent of the choice of Cauchy surface,
assuming the appropriate boundary conditions. A choice of vacuum state determines a
positive-frequency decomposition of the classical
solution space. More precisely, it specifies a
one-particle Hilbert space $\mathcal{H}_{\mathrm{in}}$
together with a real-linear map $ K:\mathscr{S}
    \longrightarrow
    \mathcal{H}_{\mathrm{in}}$
which associates each real classical solution
with its positive-frequency part. The inner
product on $\mathcal{H}_{\mathrm{in}}$ is induced
by the symplectic structure. In particular,
\begin{equation}
    2\,\mathrm{Im}
    \langle KA^{(1)},KA^{(2)}\rangle
    =
    \Omega(A^{(1)},A^{(2)}),
    \label{eq:one_particle_symplectic}
\end{equation}
The corresponding bosonic Fock space is $ \mathcal{F}(\mathcal{H}_{\mathrm{in}})
    =
    \bigoplus_{n=0}^{\infty}
    \mathrm{Sym}^n
    \mathcal{H}_{\mathrm{in}}$ with the corresponding vacuum state $|\Omega\rangle$.  Now, for a classical solution $A\in\mathscr{S}$, the associated normalized coherent state is
\begin{equation}
    |\Phi_A\rangle
    =
    \exp\left(
        -\frac{1}{2}\|KA\|^2
    \right)
    \exp\left(
        a^\dagger(KA)
    \right)
    |\Omega\rangle,
    \label{eq:EM_coherent}
\end{equation}
where $a^\dagger(KA)$ denotes the creation
operator associated with the one-particle
state $KA$. In the DSW setting, the electromagnetic field state $\ket{\Phi_i}$ associated with each branch $i=1,2$ is described by a classical source-free solution $\mathcal{A}^{(i)}$ and the corresponding quantum states are coherent states
\begin{align}
    \ket{\Phi_i} = \exp \bigg( - \frac{1}{2} \vert \vert K \mathcal{A}^{(i)} \vert \vert^2 \bigg) \exp (a^{\dagger} (K \mathcal{A}^{(i)})) \ket{\Omega}
\end{align}
We also denote the difference of these radiation solutions as $\Delta A= \mathcal{A}^{(1)}- \mathcal{A}^{(2)}$. Using Backer-Campbell-Hausdorff formula and the commutation relation $[a(f), a^{\dagger} (g) ]= \langle f, g \rangle, \: f,g \in \mathcal{H}_{\mathrm{in}}$, we can evaluate the overlap of the two coherent states to be
\begin{align}
    \bra{\Phi_2} \Phi_1 \rangle= \exp \bigg[- \frac{1}{2}\vert \vert K \mathcal{A}^{(1)} \vert \vert ^2 - \frac{1}{2} K \mathcal{A}^{(2)} \vert \vert + \langle K \mathcal{A}^{(2)}, K \mathcal{A}^{(1)} \rangle  \bigg]= \exp \bigg[- \frac{1}{2} \vert \vert K \Delta A \vert \vert^2 \bigg]
\end{align}
We interpret $\vert \vert K \Delta A \vert \vert^2$ as the expectation value of the number of entangling photons $N_{\mathrm{ent}}$ associated with the difference between the two electromagnetic radiation states
\begin{align}
    N_{\mathrm{ent}}:= \vert \vert K \Delta A \vert \vert ^2 
\end{align}
Therefore, the coherence factor is given by 
\begin{align}
    \vert \langle{\Phi_2} \vert \Phi_1 \rangle \vert = e^{- \frac{N_{\mathrm{ent}}}{2}}
\end{align}

After the holding period, the two wave packets are
coherently recombined by reversing the splitting
procedure. The degree of coherence $\vert \langle{\Phi_2} \vert \Phi_1 \rangle \vert$ can then be probed
through some kind of interference experiment. This is basically the whole setup of the DSW Gedanken experiment.
Theoretically, the remaining task is to calculate the one particle norm of the classical retarded solution $\Delta A$. For the remainder of this subsection, we specialize to a concrete example : the Rindler horizon in Minkowski space to illustrate the idea of DSW.

\par We first  outline the strategy of the DSW calculation. The two branches of the superposition are modeled as follows: in the first branch, the charged particle is held stationary at a fixed Rindler coordinate $\rho$, while in the second branch, it is displaced by a small distance $d$ in the $\rho$ direction. Each stationary
charge configuration generates a retarded electromagnetic
field whose gauge potential can be restricted to the
future Rindler horizon. Since the charged particle remains
outside the horizon, no charged matter crosses it, and
the electromagnetic field satisfies the source-free
Maxwell equations in a neighborhood of the horizon.
The retarded solutions, together with the corresponding
boundary conditions, determine the gauge potentials
induced on the horizon by the two charge configurations.
Their difference, denoted by $\Delta A$, characterizes
the electromagnetic field perturbation associated with
the displacement of the charged particle. The next step is to evaluate the one-particle norm
$\|K\Delta A\|^2$ using the symplectic structure on
the horizon and the positive-frequency decomposition
with respect to the affine null coordinate $V$. 

\par 
We begin by recalling the retarded gauge potential
generated by a point charge in Minkowski spacetime.
In suitable gauge, Maxwell's equations reduce to
$\Box A^\mu=-j^\mu$, whose retarded solution is
\begin{equation}
    A^\mu(x)
    =
    \int d^4x'\,
    G_{\mathrm{ret}}(x,x')j^\mu(x').
\end{equation}
For a point charge $q$ moving along a worldline
$z^\mu(\tau)$ with four-velocity
$w^\mu(\tau)= \frac{dz^\mu}{d\tau}$, the current is given by
\begin{equation}
    j^\mu(x)
    =
    q\int d\tau\,
    w^\mu(\tau)
    \delta^{(4)}(x-z(\tau)).
\end{equation}
Substituting this current into the retarded solution
and performing the integration gives the familiar retarded potential
\begin{equation}
    A^\mu(x)
    =
    \left.
    \frac{q\,w^\mu(\tau)}
    {4\pi[-R_\nu w^\nu(\tau)]}
    \right|_{\tau=\tau_{\mathrm{ret}}}
    \label{eq:LW_potential}
\end{equation}
where $ R^\mu=x^\mu-z^\mu(\tau_{\mathrm{ret}})$. The retarded proper time $\tau_{\mathrm{ret}}$ is determined by the condition that $R^{\mu}$ is lightlike, i.e., $R^\mu R_\mu=0$. The spacetime point $x^{\mu}$ lies in the causal future of $z^{\mu} (\tau_{\mathrm{ret}})$. We now specialize to a uniformly accelerated charge (or s stationary charge at fixed Rindler coordinate $\rho$)
in Minkowski spacetime. Let $(t,x,y,z)$ be standard
inertial coordinates, in which the metric takes the
form $ds^2=-dt^2+dx^2+dy^2+dz^2$.  To introduce coordinates on Rindler wedge, we let $t= \rho \sinh \eta$  and $x= \rho \cosh \eta$. So, the metric for the Rindler wedge is $ds^2= -\rho^2 d \eta^2 + d \rho^2 + dy^2 + dz^2$. We recall that a stationary trajectory at $\rho=D$ has proper acceleration $a= \frac{1}{\rho}$. Its proper time $\tau= \rho \eta = \frac{1}{a}\eta$. 

\par We model two branches of superposition as follows: One branch corresponds to a stationary charge at $\rho=D$ while the other branch corresponds to displacing a stationary charge by a small distance $d \ll D$ so that it is at $\rho=D+ d$. The branches initially coincide, separate smoothly, remain separated for a Killing time interval $T$ and then recombine smoothly. We call $T$ to be the holding time. Now, we introduce the null coordinates:
\begin{align}
    U = \frac{t-x}{2}, \quad V= t+x
\end{align}
so that the metric in this coordinates is
\begin{align}
    g = -2 dU dV + dy^2 + dz^2 \label{Rindler UV metric}
\end{align}
Note that (\ref{Rindler UV metric}) takes the form of (\ref{eq:bifurcate_metric}) with $A=1$ and $h_{ab}= \delta_{ab}$. The Rindler horizon $\Sigma_+$ is located at $U=0$ and the Killing vector field restricted  to $\Sigma_{+}$ is simply $\xi = V \partial_V$. We also note that since $V= t+x = \rho e^{\eta}$, we have the following relation
\begin{align}
    \log V= \eta + \log \rho  \label{IMMM logarithmic relation}
\end{align}
So, the Killing time $\eta$ is in the logarithmic relation with the null cooridinate $V$ up to a radial $\rho$ dependent term in the Rindler wedge. Now, we want to calculate  the retarded gauge field  $A_{\mu}$ due to the stationary charge at $\rho= D$. We will perform this calculation in null coordinates $x^{\mu}= (U,V,y,z)$.  The worldline trajectory of the  stationary charge at $\rho= D$ in this coordinate system is \footnote{In Cartesian coordinates, we let the worldline of the stationary charge at $\rho=D$ to be $z^{\mu}(\eta)=(D \sinh \eta, D\cosh \eta,0,0)$.}
\begin{align}
    z^{\mu} (p)= (U=- \frac{D}{2}e^{- \eta}, V= De^{\eta}, 0,0)= (- \frac{D}{2p}, Dp, 0,0)
\end{align}
where we let $p= e^{\eta}$ (note that $p$ is not free for our purpose of calculating the retarded gauge field $A_{\mu}$, it is constrained by the lightlike condition of the separation retarded source point $R^{\mu}$). Its $4$-velocity $w^{\mu}= \frac{dz^{\mu}}{ d \tau}= \frac{1}{D} \frac{dz^{\mu}}{d \eta}$ is 
\begin{align}
    w^{\mu}= (\frac{1}{2p},p,0,0)
\end{align}
Now, we denote a spacetime point $x^{\mu}= (U,V,y,z)$ which is to the causal future of $z^{\mu} (p)$. Its separation from the retarded source point is  $R^{\mu}= x^{\mu}- z^{\mu}(p)$. Its norm square is
\begin{align}
    R_{\mu}R^{\mu}= -2 UV + 2 U Dp -\frac{DV}{p} + D^2 + r^2 =0
\end{align}
where we let $y^2 + z^2 = r^2$.  We also calculate the retarded distance $\mathfrak{R}= - R^{\mu}w_{\mu}$ 
\begin{align}
    \mathfrak{R}= Up + \frac{V}{2p}
\end{align}
So, the retarded gauge field $A_{\mu} (x)= \frac{q w_{\mu}}{4 \pi \mathfrak{R}}$ is
\begin{align}
    A_{\mu} (x)= \bigg( \frac{-q p^2}{2 \pi (V+2Up^2)},  \frac{-q}{4 \pi (V+ 2U)}, 0,0 \bigg)
\end{align}
For later purpose, we also note that $ \partial_U A_V = \frac{q}{4 \pi}\frac{2p^2 + 4 Up\: \partial_U p}{(V+ 2Up^2)^2} $. Now, we let $X^{\mu} =(0, V,y,z)$ to be a spacetime point on the Rindler horizon $\Sigma_+$. The corresponding retarded gauge field $A_{\mu} (X)$ is simply (let $U=0$)
\begin{align}
    A_{\mu} (X)= \bigg( \frac{-qp^2}{2 \pi V}, \frac{-q}{4 \pi V},0,0  \bigg)
\end{align}
and the corresponding $p (X)$ is given by
\begin{align}
    p(X)= \frac{DV}{D^2 + r^2}
\end{align}
So, $\partial_U A_V \vert_{\Sigma_=}= \frac{qp^2}{2 \pi V^2}=\frac{qD^2}{2 \pi (D^2 + r^2)^2}$. Since $\partial_V$ is tangent to the Rindler horizon $\Sigma_+$, we also have $\partial_{V}A_{U}\vert_{\Sigma_+}= \partial_V A_{U} (X)=- \frac{q D^2}{2 \pi (D^2 + r^2)^2}$. Therefore, the Coulomb field  $C_{D} (r):=F_{UV}\vert_{\Sigma_+}= \partial_U A_{V} \vert_{\Sigma_+}- \partial_V A_{U} \vert_{\Sigma_+}$ along the horizon is 
\begin{align}
    C_D (r)= \frac{q D^2}{\pi (D^2 + r^2)^2}
\end{align}
For this stationary configuration and along the horizon $\Sigma_+$, the electric field $E_{A}= F_{AV}= \partial_A A_V=0, \: A=y,z$. So,  there is no electric field in the transverse direction ($E_A =0$)  on the horizon for this case.

\par Now, we want to discuss the effect due to displacing the stationary charge by a small distance $d$ with $\vert \vert d \vert \vert \ll D$. The change of the Coulomb field due to such  small displacement $d$ is given by
\begin{align}
    \Delta C &= C_{D+d}(r)- D_{D}(r)= d \: \partial_D C_{D} + \mathcal{O} (d^2) \\
    &= \frac{2qd D (r^2- D^2)}{\pi (D^2 + r^2)^3} + \mathcal{O} (d^2)  \label{Delta C eqn}
\end{align}
We will let $\Delta C \simeq   \frac{2qd D (r^2- D^2)}{\pi (D^2 + r^2)^3} $ and ignore the subleading corrections in $d$. We next determine what Maxwell’s equations on the horizon $\Sigma_+$ that require during the transition between them. The charged particle remains outside the horizon, so  no charged current crosses the horizon. Maxwell’s equation there is $\partial_a F^{aU}=0$. Identifying $F^{VU}=C, F^{AU}=- E^{A}$, we have
\begin{align}
    \partial_A E^{A}= \partial_V C, \quad A=y,z
\end{align}
So, the change of the electric fields $E^A$ due to the displacement of charge is governed by the equation $\partial_A \Delta E^A = \partial_V \Delta C$.  Now, we perform a further gauge transformation or fix a guage such that $A_{V}=0$. It follows that 
\begin{align}
  \partial^A \Delta A_A=- \Delta C
\end{align}
For our axisymmetric displacement, , we can also write in polar coordinates that underly $\Delta A_A= \Delta A (r) \hat{r}_A$ (recall that $r^2=x^2 + y^2$). Then, we have $\partial^A \Delta A_A= \frac{1}{r} \frac{d}{dr} (r \Delta A(r))$. It follows that $r \Delta A (r)=- \int^{r}_0 ds \: s  \Delta C(s)$. Substituting (\ref{Delta C eqn}) into this equation and perform the integration, we obtain
\begin{align}
    \Delta A(r)= \frac{qdDr}{\pi (D^2+r^2)^2}
\end{align}
We also define the spatial strength of the memory $S_M$ as
\begin{align}
    S_M =  \int d^2 x  \: \Delta A_A \Delta A^{A}
\end{align}
Straightforward calculation gives
\begin{align}
    S_M = \frac{2q^2d^2 D^2}{\pi} \int^{\infty}_0 dr \: \frac{r^3}{(D^2 + r^2)^4}= \frac{q^2 d^2}{6 \pi D^2}
\end{align}
Now, we want to convert the horizon potential into a one particle Hilbert space norm. For radiative horizon data with gauge $A_{V}=0$, the symplectic form $\Omega_{\Sigma_{+}}$ becomes 
\begin{align}
    \Omega_{\Sigma_{+}} (A,B) = \int dV d^2 x \big( A_A \partial_V B^A -B_A \partial_V A^A   \big)
\end{align}
We can define the Fourier transform $\hat{A}_A  (\omega,x)= \int dV e^{i \omega V} A_A (V,x)$. The positive frequency part of $A_A (V,x)$ is $ A^{+}_{A} (V,x)=  \int^{\infty}_{0} \frac{d \omega}{2 \pi} e^{-i \omega V} \hat{A}_A (\omega,x)$. Its norm is given by $\vert \vert A^{+}_{A} \vert \vert^2= i \Omega_{\Sigma_+} (\bar{A}^+_A,A^{+}_A )$. It follows that the horizon contribution to the difference-field occupation number of photons $N_{\Sigma_+}$ is given by
\begin{align}
    N_{\Sigma_+}= \frac{1}{ \pi} \int d^2 x \int_{0}^{\infty} d \omega \: \omega \vert {\widehat{\delta A}_A (\omega,x)} \vert^2
\end{align}
where $\delta A_A (V,x)= \Delta A_A (x) f(V)$. Here, we introduce an extra function $f(V)$ because $\Delta A (x)$ only tells us  the change between the initial and final
stationary horizon potentials in a gauge fixed consistently throughout
the displacement but it does not yet specify how the potential evolves
during the transition. To calculate its Fourier transform, and hence
its one-particle norm, we must also take into account this time dependence $f(V)$. This function $f(V)$ describes the progress of
the displacement along affine horizon time. We first consider a displacement that is not subsequently reversed.
The two branch currents initially agree. After the displacement,
one branch remains at the original position, while the other remains
at the displaced position. Accordingly, we choose a smooth function that rises from zero to one i.e., $f(V)=0$ when $V \leq V_i$ and $f(V)=1$ when  $V  \geq V_f $. We define the distribution $g(V)= f'(V)$. Its Fourier transform is
\begin{align}
    \hat{g}(\omega)= \int^{V_f}_{V_i} dV e^{i \omega V} g(V)=-i  \omega \hat{f}(\omega)
\end{align}
Since $g(V)=0$ outside the transition interval, $\hat{g}(\omega)$ is a convergent integral. Hence, away from low frequency region, we have $ \hat{f}(\omega)= \frac{i}{\omega} \hat{g} (\omega)$. Therefore, it leads to that $N_{\Sigma_+} \sim \frac{S_M}{\pi} \int_0 \frac{1}{\omega} $ which is  a logarithmic infrared divergence. This permanent-memory limit is not an ordinary finite-norm coherent state. For the DSW experiment, we should instead calculate a finite splitting-and-recombination protocol. 

\par Now, we let $f$ slowly rises to one, remains approximately constant, and returns to zero, During displacement, $f'(V) >0$; during the holding period, $f'(V) \approx 0$ and during recombination, $f'(V) <0$. We can define a positive displacemnt pulse, i.e, $p_{disp}=f'(V)$ during splitting and $p_{rec}=0$ elsewhere. Similarly, we can define a positive recombination pulse, i.e, $p_{rec} (V) \sim -p _{disp}(\frac{V}{R})$ where $R= e^{\frac{T}{D}}$, $T$ is the elapsed proper holding time of the stationary charge or the laboratory where the experiment is performed. Note that on the horizon $\Sigma_{+},$ the affine null time is $V(T)= V_0e^{\frac{T}{D}}$. We let the displacement time $V_{disp}=V_0$ and the recombination time to be $V_{rec}= V_0 e^{\frac{T}{D}}$. Therefore, the recombination pulse $-p_{rec}(V)$ is just the displacement pulse $p_{disp}$ with the arguments shifted by $R$. Hence, we have
\begin{align}
    f'(V)=p_{disp} (V)- \frac{1}{R}p_{disp} \big(\frac{V}{R} \big)  \label{f'V pulse}
\end{align}
We required that the pulse $p_{disp}$ is normalized, i.e,
\begin{align}
    \int_{- \infty}^{\infty} dV \: p_{disp} (V)=1
\end{align}
Note that a factor $\frac{1}{R}$ is included in the second term of (\ref{f'V pulse}) so that the recombination pulse is also normalized. The corresponding Fourier transform is
\begin{align}
    \widehat{f'} (\omega)= \widehat{p}_{disp} (\omega)- \hat{p}_{disp} (R \omega)
\end{align}
To proceed the calculation, we let $\omega= \omega_0 e^{s}$ where $\omega_0 >0$ and we define $Q (s)= \hat{p}_{disp} (\omega)= \hat{p}_{disp} (\omega_0e^s)$. Since $\hat{p}_{disp} (0)=1$ and $\hat{p}_{disp} (\infty)=0$, it follows that $Q (- \infty)=1$ and $Q (\infty)=0$. So, $Q$ comes from the left as $1$ and comes from the right as $0$. With respect to this new variable $s$ with the range $- \infty < s < \infty$,  we have
\begin{align}
    \widehat{f'} (s)= Q(s)- Q (s + \log R) 
\end{align}
Consequently, the difference-field occupation number of photons $N_{\Sigma_+}$ is (use $\hat{f}(\omega)= \frac{1}{\omega}\widehat{f'}(\omega)$ and $ \frac{d \omega}{ \omega}= ds$)
\begin{align}
    N_{\Sigma_+} = \frac{S_M}{\pi} \int_{- \infty}^{\infty} ds \: \vert Q(s)- Q(s + \log R) \vert^2
\end{align}
Now, we approximate $Q(s)$ as a sharp step function $Q_0 (s)= \begin{cases}
    1 & \text{if } s < 0 \\
    0 & \text{if } s > 0 
  \end{cases} $ . Then, we have 
$Q_0 (s)- Q_0 (s + \log R)= \begin{cases}
    1 & \text{if } - \log R<s < 0 \\
    0 & \text{elsewhere }  
  \end{cases} $ . So, we know that $ \int ds\vert Q_0 (s)- Q_{0} (s + \log R) \vert^2= \log R$. Certainly for the smooth pulse, $Q$ is not a sharp transition and can be more complex, but its departure from the sharp transition is controllable  in an integrable sense. To see this, we let
  \begin{align}
      Q(s)= Q_{0} (s) + r(s)
  \end{align}
where $r(s)$ is a perturbation that represents the depature from sharp transition. We will denote $G(s)=Q_0 (s)- Q_0 (s + \log R)$ and $Z(s)= r(s)- r(s + \log R)$. Then , we have
\begin{align}
    Q(s)- Q(s + \log R)= G(s)+ Z(s)
\end{align}
We already know that $\int_{- \infty}^{\infty} ds \: \vert G(s) \vert^2 =\log R$. Using $\vert G +Z \vert^2= \vert G \vert^2 + \vert Z \vert^2 + 2 \Re (\bar{G}Z)$, we know that
\begin{align}
    \int^{\infty}_{- \infty} ds \; \vert Q(x)- Q (x + \log R) \vert^2 = \log R + C_{R}
\end{align}
where $C_{R}$ denotes correction which is
\begin{align}
    C_R = 2 \Re \int^{\infty}_{- \infty}  ds\overline{G(s)} Z (s) + \int^{\infty}_{- \infty} ds \vert Z(s) \vert^2  \label{correction in CR}
\end{align}
For the first term in (\ref{correction in CR}), we have
\begin{align*}
    \vert 2 \Re \int^{\infty}_{- \infty}  ds\overline{G(s)} Z (s) \vert &\leq 2  \int^{\infty}_{- \infty} ds \vert G(s) \vert \cdot\vert{Z(s)} \vert \\ 
    &\leq 2 \int^{\infty}_{- \infty} ds \vert Z (s) \vert \\
    & \leq 2 \int^{\infty}_{- \infty} ds  \big( r(s) + r(s+ \log R)   \big)\\
    &= 4 \vert \vert r \vert \vert_{L^1}
\end{align*}
In the second inequality, we use $\vert G(s) \vert \leq 1$. Moreover, for the second term in (\ref{correction in CR}), we can also do an estimate
\begin{align*}
    \int^{\infty}_{- \infty} ds \: \vert Z(s) \vert^2 \leq  2 \int^{\infty}_{- \infty} ds \; \big( \vert r(s) \vert^2 + \vert r(s + \log R) \vert^2  \big)= 4 \vert \vert r \vert \vert_{L^2}
\end{align*}
where we have used $\vert a- b \vert^2  \leq  2 \vert a \vert^2 + 2 \vert b \vert^2$. Therefore, we see that if $r(s) \in L^{1} (\mathbb{R}) \cap L^2 (\mathbb{R})$, we have
\begin{align}
\Bigg| \int^{\infty}_{- \infty} ds \vert Q(x)- Q(x + \log R ) \vert^2- \log R \Bigg| \leq 4 \big( \vert \vert r \vert \vert_{L^1} +  \vert \vert r \vert \vert_{L^2} \big)  \label{shape correction}
\end{align}
Hence, we know that
\begin{align}
    N_{\Sigma_+} = \frac{S_M}{ \pi} \log R + \mathcal{O}(1) 
    \approx \frac{q^2 d^2}{6 \pi^2D^3} T
\end{align}
In particular,  from (\ref{shape correction}), we know that the correction $C_R$ only depends on the shape of the pulse and does not affect the leading behaviour of $N_{\Sigma _+} $ which is linear in the holding proper time $T$. So, we obtain 

\begin{align}
    \vert \langle \Phi_2 \vert \Phi_1 \rangle \vert \approx  \exp \bigg \{- \frac{q^2 d^2 T}{12 \pi^2 D^3} \bigg \}
\end{align}
In this case, the dephasing rate is
$\Gamma_{\mathrm{Rindler}}=\frac{q^2d^2}{12\pi^2D^3}$.
As shown by DSW \cite{Danielson:2022sga}, the exponential behavior
$\vert\langle\Phi_2\vert\Phi_1\rangle\vert\approx \exp(-\Gamma T)$,
where $T$ is the elapsed proper holding time of the laboratory,
holds generally in spacetimes with Killing horizons admitting
a metric of the form~(\ref{eq:bifurcate_metric}).
The underlying physics is entirely analogous to that of the
example demonstrated above. We thus conclude that Killing horizons
decohere quantum superpositions.

\section{ETH Approach to Horizon-Induced Decoherence}
In this section, we  demonstrate how the essential features of horizon-induced decoherence can be understood within the framework of the Fr\"ohlich-Pizzo (FP) model. As the original discrete construction is not directly suited to the continuum nature of horizon quantum fields, we introduce an adapted version that preserves its underlying physical mechanism. Before introducing our adapted FP model, we will give a general motivation first.

\subsection{Motivation}
\par We recall that a key feature of the FP model is the
Principle of Diminishing Potentialities (PDP), manifested
in the strict inclusion $ \mathcal{E}_{\geq \tau'}
    \subsetneq
    \mathcal{E}_{\geq \tau},
    \: \tau' > \tau$
which originates from the corresponding shrinking of
the algebras associated with the $R$-field system $\mathcal{B}_{\geq \tau'}
    \subsetneq
    \mathcal{B}_{\geq \tau},
    \: \tau' > \tau$.
 A closely related algebraic structure
arises naturally in quantum field theory on Killing
horizons, where the algebras of observables associated
with successive horizon cuts form a decreasing family. To make this connection precise, we first need to
make clear the notion of a horizon-cut algebra. It is
well known that restricting quantum fields to a null
hypersurface presents certain mathematical subtleties.
In ordinary QFT, local fields are operator-valued
distributions, and smearing them solely along a null
hypersurface does not, in general, yield well-defined
operators on the Hilbert space $\mathcal{H}$.\footnote{
An an example, we can consider a local operator
$\mathcal{O}(x)$ with  scaling dimension
$\Delta$, whose two-point function behaves schematically
as
$\langle \mathcal{O}^{\dagger}(x)
\mathcal{O}(x')\rangle
\sim [(x-x')^2]^{-\Delta}$
at short distances.
In null coordinates $(U,V,y^A)$, the invariant
separation takes the form
$(x-x')^2=-(U-U')(V-V')+|y-y'|^2$.
Restricting both points to the null hypersurface
$U=0$ formally gives
$(x-x')^2=|y-y'|^2$.
Consequently, the squared norm of a horizon-smeared
field formally involves a transverse integral of
the form
\begin{align}
    \int_{|r|<\epsilon}
    \frac{d^{d-2}r}{|r|^{2\Delta}}
    \propto
    \int_0^\epsilon dr\,
    r^{d-3-2\Delta},
\end{align}
which diverges for $2\Delta\geq d-2$.}
Nevertheless, matrix elements of local fields $\langle\psi|\mathcal{O}(x)|\phi\rangle,
    \: x\in\Sigma_+$
may remain meaningful as sesquilinear forms. We recall that a sesquilinear form
on a dense linear subspace
$\mathcal{D}\subset\mathcal{H}$
is a map
\begin{align}
    \mathfrak{s}:
    \mathcal{D}\times\mathcal{D}
    \longrightarrow \mathbb{C},
\end{align}
which is conjugate-linear in its first argument and
linear in its second. A local
field $\mathcal{O}(x)$, with $x\in\Sigma_+$, may
define such a form through its matrix elements
\begin{align}
    \mathfrak{s}_{\mathcal{O}(x)}
    (\psi,\phi)
    :=
    \langle\psi|\mathcal{O}(x)|\phi\rangle,
    \qquad \psi,\phi\in\mathcal{D}.
\end{align}
A modern approach, discussed in
\cite{Faulkner:2024gst}, is therefore to begin with
these sesquilinear forms and construct the associated
von Neumann algebra through their double commutant. The procedure goes as follows. First, we consider the future horizon $\Sigma_+$ equipped with
an affine parameter $V$ along its null generators and
transverse coordinates $x^A$.  A horizon cut is specified
by $V=\lambda$, and we denote its future portion by
\begin{equation}
    \Sigma_\lambda^+
    =
    \{(V,x^A)\in \Sigma_+ \vert 
    V \geq\lambda\}.
\end{equation}
We denote $\mathcal{W}_{\lambda}$ as a collection of densely defined
sesquilinear forms $\mathfrak{s}_{\mathcal{O}(x)}$ associated with local quantum fields
on $\Sigma_\lambda^+$. The domain $\mathcal{D}_{\mathfrak{s}_{\mathcal{O}(x)}}$ associated to a given local operator $\mathcal{O}(x)$ viewed as a sesquilinear norm is also expected to be dense in $\mathcal{H}$ (see section $3.1$ of \cite{Faulkner:2024gst}). Now, a bounded operator $B$ is said to commute with a sesquilinear form $\mathfrak{s}$ if $B$ preserves the domain $\mathcal{D}_{\mathfrak{s}}$ of $\mathfrak{s}$ and if it satisfies
\begin{align}
    \mathfrak{s} (B^* \psi, \phi) = \mathfrak{s} (\psi, B \phi)  \label{commutatnt criteria}
\end{align}
 The motivation for this criteria is straightforward, since if the sesquilinear form is induced by the local operator $\mathcal{O} (x)$, then (\ref{commutatnt criteria}) just means that $B$ commutes with $\mathcal{O}(x)$ in matrix elements, i.e,
 \begin{align}
     \mathfrak{s}_{\mathcal{O}(x)} (B^* \psi, \phi)= \langle \psi \vert B \mathcal{O} (x) \vert\phi \rangle= \langle \psi  \vert \mathcal{O}(x)B \vert \phi \rangle= \mathfrak{s}_{\mathcal{O}(x) } (\psi, B \phi), \quad \psi, \phi \in \mathcal{D}_{\mathfrak{s}_{\mathcal{O} (x)}}
 \end{align}
 The commutant of $\mathcal{W}_{\lambda}$, denoted by
$\mathcal{W}_{\lambda}'$, is identified with the
algebra of bounded operators localized in the left
wedge $W'_{\lambda}$, which is spacelike separated from the portion
of the horizon to the future of the cut $V=\lambda$. See Figure 1a. We then  define the full horizon cut algebra $\mathcal{M}_{\geq \lambda}$ associated to the region $\Sigma^{+}_{\lambda}$ by
\begin{align}
    \mathcal{M}_{\geq \lambda} :=  (\mathcal{W}_{\lambda}')'=\mathcal{W}_{\lambda}''
\end{align}
 By construction, this algebra can be identified with the algebra of  bounded operators localized in the region $W_{\lambda} \cup \Sigma^{+}_{\lambda}$ as illustrated in Figure 1a. Furthermore, for two horizon cuts satisfying $\lambda' > \lambda$,  the corresponding regions obey $W_{\lambda'} \subseteq W_{\lambda}$ and $\Sigma^{+}_{\lambda'} \subseteq \Sigma^{+}_{\lambda}$. By isotony, this implies the inclusion for horizon cut algebras $\mathcal{M}_{\geq \lambda' } \subseteq \mathcal{M}_{\geq \lambda} ; \: \lambda' > \lambda$. It is also known that this nested structure of horizon-cut algebras play the central role  for the generalized second law $\Delta \mathcal{S}_{gen} \geq 0$ to hold true, as first shown in \cite{Wall:2011hj}, where the monotonicity of relative entropy under restriction to smaller algebras provides a key ingredient of the proof. In fact, by a simple analysis as shown in section $4.1$ of \cite{Sia:2025rux}, this inclusion of horizon cut algebras is also expected to be strict, i.e.,
 \begin{align}
     \mathcal{M}_{\geq \lambda'} \subsetneq \mathcal{M}_{\geq \lambda}, \quad \lambda' > \lambda  \label{PDP NULL ALGEBRAS}
 \end{align}
Thus, the family of horizon-cut algebras exhibits PDP, suggesting a natural connection with the FP model. In this correspondence, the horizon-cut algebras $\mathcal{M}_{\geq\lambda}$ play a role analogous to that of the $R$-field algebras $\mathcal{B}_{\geq\tau}$ in the FP model, as both form decreasing families of von Neumann algebras encoding the progressive restriction of future potentialities. Meanwhile, the charged particle plays the role of the ``atom'' in the FP model. Just as the interaction between the atom and the $R$-fields generates entanglement characterized by a Gram matrix, the charged particle becomes entangled with the electromagnetic field through the branch-dependent retarded field reaching the horizon. The corresponding Gram matrix quantifies the overlap between the horizon field states associated with different particle branches, thereby governing the suppression of quantum coherence between them. This structural correspondence motivates us to introduce an adapted FP model that preserves the essential features of the original discrete construction while incorporating the causal structure of a Killing horizon.

\begin{figure}[htbp]
    \centering
    \includegraphics[width=0.8\textwidth]{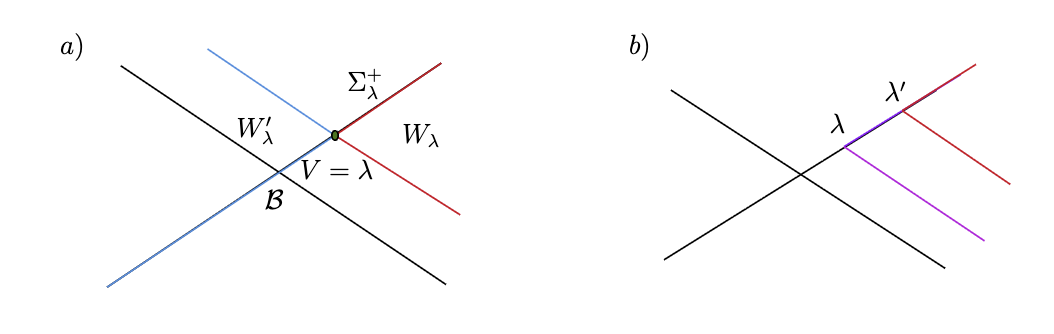}
    \caption{ Schematic illustration of the horizon cuts and their
associated wedge regions.
(a) The bifurcation surface is denoted by $\mathcal{B}$,
and the green dot represents the codimension-two
cross-section $V=\lambda$ of the future horizon.
The portion of the horizon to the future of this cut
is denoted by $\Sigma^+_{\lambda}$ (red).
The red and blue lines bound the right wedge
$W_{\lambda}$ and its causal complement
$W'_{\lambda}$, respectively.
(b) Two horizon cuts with $\lambda'>\lambda$.
The corresponding wedge regions are nested,
$W_{\lambda'}\subset W_{\lambda}$,
as are the future portions of the horizon,
$\Sigma^+_{\lambda'}\subset\Sigma^+_{\lambda}$.}
    \label{fig:figure1}
\end{figure}

\subsection{Adapted FP Model for Horizon Decoherence}
\subsubsection{$U(1)$ chiral current for horizon's degree of freedom}
The principal modification in our adaptation of the FP model is to replace the infinite tensor product of matrix algebras describing the $R$-field system with the local von Neumann algebras of the chiral $U(1)$ current \cite{Buchholz:1988ae}. As we shall see, this adaptation preserves the underlying structure of the original FP model while accommodating the causal structure of Killing horizons \cite{1993CMaPh.156..201B} \cite{Borchers:1998ye}. To describe horizon cut algebras in our model, we first consider massless scalar field in $1+1$ dimensional flat space
\begin{align}
    S[\phi]= \frac{1}{2} \int dt \: dx \big( (\partial_t \phi)^2 - (\partial_x \phi)^2  \big)  \label{massless action}
\end{align}
We can introduce null coordinates $U= t+x, V= t-x$. The equation of motion for  (\ref{massless action}) is
\begin{align}
    \partial_V \partial_U \phi=0
\end{align}
Therefore, the solution is 
\begin{align}
    \phi= \phi_R (U) + \phi_L (V)
\end{align}
In our model, we will only keep the left moving mode $\phi_L (V)$ and discard the right moving mode $\phi_R (U)$. The left moving mode $\phi_L (V)$ admits the standard Fourier decomposition in fequency space $\omega$
\begin{align}
    \phi_L (V) = \int^{\infty}_{0} \frac{d \omega}{\sqrt{2 \pi \omega}}  \bigg( a_\omega e^{-i \omega V} + a^{\dagger}_{\omega} e^{i \omega V}   \bigg)
\end{align}
The corresponding vaccum state is characterized by $a(\omega) \ket{0}=0, \forall \omega>0$. The corresponding $V$-translation generator is given by
\begin{align}
    P_V= \int^{\infty}_{0} d \omega  \: \omega \: a^{\dagger}_\omega a_\omega
\end{align}
and it annihilates the vaccum state $P_V \ket{0}=0$. It is not feasible to directly work with left moving mode $\phi_L (V)$ due to the infrared zero modes problem. In particular, its two point function behaves as
\begin{align}
    \omega_0 (\phi_L (V) \phi_L (V')):= \langle 0 \vert \phi_L (V) \phi_L (V') \vert 0 \rangle = \int^{\infty}_{0} \frac{d \omega}{2 \pi} \frac{1}{\omega} \: e^{-i \omega (V-V')}
\end{align}
Around the frequency $\omega \rightarrow 0$, it contains a logarithmic IR divergence part $\int^{\epsilon}_{0} \frac{d \omega}{\omega} \rightarrow \infty$. This leads to that the smeared field $\phi_L (f)= \int^{\infty}_{- \infty} dV \: f(V) \phi_L (V), \: f \in C^{\infty}_{c}(\mathbb{R})$ is not a well defined operator, i.e.,
\begin{align}
    \omega_0 (\phi_L (f)^2)= \langle 0 \vert \phi_L (f)^2 \vert 0 \rangle = \int^{\infty}_0 \frac{d \omega}{2 \pi \omega} \vert \hat{f} (\omega) \vert^2
\end{align}
Since $\hat{f}(0)= \int^{\infty}_{- \infty} d V f(V)$, so in general $\hat{f}(0) \neq 0$ and the integral will still diverges. Hence, we see that the  usual massless vacuum cannot be defined as a regular Gaussian state on the full algebra of these smeared chiral scalar fields. So, instead, we define a $U(1)$ current 
\begin{align}
    J(V)=
 \partial_V \phi_L (V)
\end{align}
Its mode expansion is
\begin{align}
    J(V)= i \int^{\infty}_0 d \omega \sqrt{\frac{\omega}{2 \pi}} \bigg( -a_ \omega e^{-i \omega V} + a^{\dagger}_{\omega} e^{i \omega V}  \bigg)
\end{align}
and the corresponding two point function is
\begin{align}
    \omega_0 (J (V) J(V'))= \int^{\infty}_{0} d \omega  \frac{\omega }{2 \pi} e^{-i \omega (V-V')}   \label{current J two}
\end{align}
Unlike the scalar two-point function, whose spectral measure behaves as $d\omega/\omega$ at low frequencies, the current two-point function carries a factor of $\omega$ in the numerator. This eliminates the infrared divergence and allows the vacuum to define a well-behaved quasifree state on the smeared current. In particular, we can smear it with a test function, i.e., $J (f)= \int^{\infty}_{- \infty} d V f(V) \partial_V \phi (V)= - \int^{\infty}_{- \infty} dV f'(V) \phi(V)$, so  we have
\begin{align}
    \omega_0 (J(f)^2)= \int^{\infty}_{0} \frac{d \omega}{2 \pi \omega} \vert \hat{f'}(\omega) \vert^2   \label{J2 formula}
\end{align}
Since $ \hat{f'}(0)=\int^{\infty}_{- \infty} dV f' (V)= f(\infty)- f(- \infty)=0$, we see that this integral (\ref{J2 formula}) is finite. So, smearing the chiral current is equivalent to smear the scalar with a zero-integral test function and the chiral current $J(f)$ is a well-defined operator. Therefore, we will just work with these chiral current $J (f)$.

\par We can also directly evaluate its commutator
\begin{align}
    [J(V), J(V')]
= \frac{-i }{\pi} \int^{\infty}_{0} d \omega \: \omega \sin (\omega (V-V'))= i \delta' (V-V')
\end{align}
Therefore, the commutator for the smeared currents is
\begin{align}
    [J(f), J(g)] &= i \int dV dV' f(V) g(V') \delta' (V-V') \\
    &=i \int d V f(V) g'(V)= -i \int dV f'(V) g(V)  \label{locality condition}
\end{align}
 We can define the symplectic form $\sigma (f,g)$ on the space $C^{\infty}_{c}(\mathbb{R})$ as
\begin{align}
    \sigma (f,g)= \int dV f(V)g'(V)
\end{align}
and we have
\begin{align}
    [J(f), J(g)]= i \sigma (f,g)
\end{align}
Moreover, since the vacuum state of the chiral $U(1)$ current is quasifree, the expectation value of the Weyl operator $W(f)=e^{iJ(f)}$ takes the Gaussian form
\begin{align}
\omega_0(W(f))
=\exp\left(-\frac{1}{2} \vert|h_f \vert|_1^2\right)
=\exp\left(-\frac{1}{2}\mu(f,f)\right),
\end{align}
where $h_f=\sqrt{\frac{\omega}{2 \pi}} \hat{f} (\omega)$ denotes the one-particle wavefunction associated with the smearing function $f$, and $\mu$ is the symmetric covariance determined by the vacuum two-point function,
\begin{align}
\mu(f,g)
&:=\frac{1}{2}\omega_0\left(\{J(f),J(g)\}\right)=\operatorname{Re}\int_0^\infty
\frac{d\omega}{2\pi}
\omega\overline{\hat f(\omega)}\hat g(\omega)
\end{align}
In particular, $\mu(f,f)= \vert |h_f \vert|_1^2$, so the vacuum expectation value of a Weyl operator is completely determined by the one-particle norm of its smearing function.

Now, we define the Weyl algebra for the chiral current as
\begin{align}
    \mathcal{W}= \{W(f) := e^{i J(f)} \vert \: f \in C^{\infty}_{c} (\mathbb{R})   \}''
\end{align}
The elements $W(f)$ also satisfies Weyl-relation (use BCH formula to get it)
\begin{align}
    W(f)W(g)= e^{\frac{-i}{2} \sigma (f,g)} \: W(f+g)
\end{align}
We can also define the local subalgebra $\mathcal{W} (I)$ as
\begin{align}
    \mathcal{W} (I)= \{ W(f) \vert supp (f) \subset I   \}''
\end{align}
where $I$ is an interval. From here, the time band algebra $\mathcal{W}_{\geq V_1}$ is given by
\begin{align}
    \mathcal{W}_{\geq V_1}= \mathcal{W} ([v_1, \infty))
\end{align}
A key feature of this $U(1)$ chiral current model is \textbf{locality}. If the support of  two test functions $f$ and $g$ are disjoint, i.e., $supp (f) \cap supp (g)= \emptyset$, then from (\ref{locality condition}) and the fact that differentiation does not enlarge the support of a smooth function, we have
\begin{align}
    [W(f), W(g)]=0, \quad supp (f) \cap supp (g)= \emptyset
\end{align}

From here, we can see that PDP holds in this setup. We let $V_2 > V_1$ and we let $I=(a,b)$ to be an interval with $V_1 < a<b<V_2$. We also let $f \in C^{\infty}_c (I)$. It follows that $W(f) \in \mathcal{W}_{\geq V_1}$. Since $I$ is disjoint from $[V_2, \infty)$, we also know that $W(f) \in \mathcal{W}'_{\geq V_2}$. So, the relative commutant $\mathcal{W}_{\geq V_1} \cap \mathcal{W}'_{\geq V_2} \neq 0$. This immediately implies PDP
\begin{align}
    \mathcal{W}_{\geq V_2}  \subsetneq \mathcal{W}_{\geq V_1}, \quad V_2 > V_1
\end{align}
Therefore, in our adapted FP model, we should use the Weyl subalgebra $\mathcal{W}_{\geq V}$ to represent the horizon cut subalgebra $\mathcal{M}_{\geq \lambda}$ and the horizon degree of freedom is represented by Weyl operator $W(f)$. Similarly , the vacuum state $\omega_0$ in our model should represent the corresponding horizon vacuum (chosen with respect to the affine null time $V$ of the future horizon $\Sigma_+$).

\subsubsection{Thermal Properties of $U(1)$ Chiral Current Algebras}

Next, we observe that the vacuum state $\omega_0$ restricted to half-positive line $V >0$ (or equivalently with respect to the Killing time $v \sim \log V$) is a finite $\beta$ KMS states. For simplicity, we will set the surface gravity $\kappa=1$. We first note that to make the two point function (\ref{current J two}) well-defined, we need to introduce a positive regulator $\epsilon$
\begin{align}
    \omega_0 (J(V) J(V')) = \frac{1}{2 \pi} \lim_{\epsilon \rightarrow 0_+}\int^{\infty}_{0} d \omega \: \omega \: e^{- (\epsilon + i (V-V')) \omega}
\end{align}
Using $\int^{\infty}_{0} d \omega \: \omega e^{-a \omega}= \frac{1}{a^2}, \: \Re
a >0$, we obtain
\begin{align}
    \omega_0 (J(V)J(V'))= - \frac{1}{2 \pi} \frac{1}{(V-V' -i \epsilon)^2}
\end{align}
Now, on the positive half-line $V > 0$, we let $ V= e^v$. We want to change to coordinate $v$. Since the current $J(V)$ has conformal weight one, we have
\begin{align}
    j(v)= \frac{dV}{dv} J(V)= e^v J (e^v)
\end{align}
It leads to  
\begin{align}
    \omega_0 (j(v) j(v')) = - \frac{1}{2 \pi} \frac{e^{v + v'}}{(e^{v} - e^{v'}-i \epsilon)^2}
\end{align}
Using $e^{v} - e^{v'}=2 e^{\frac{v + v'}{2}} \sinh \big( \frac{v-v'}{2}  \big)$, we obtain
\begin{align}
    \omega_0 (j(v) j(v'))= - \frac{1}{8 \pi} \frac{1}{\sinh^2 \big(  \frac{v-v' -i \epsilon}{2} \big)}  \label{Gibbs form}
\end{align}
(\ref{Gibbs form}) depends on $v-v'$. This also implies that the vacuum restricted to $V>0$ is invariant under dilation $V \mapsto e^s V$. To verify KMS condition, we let $G^+ (v)= \langle j(V) j(0) \rangle_0= - \frac{1}{8 \pi} \frac{1}{\sinh^2 (v/2)}$. Now, we introduce a complex function $F(z)= G^{+}(z)$ by analytic continuation. The singularities of $F(z)$ are at $z = 2 \pi i n, n \in \mathbb{Z}$. So, $F(z)$ is analytic in the open strip $- 2 \pi < \Im{z}<0$. Furthermore, since $\sinh (\frac{z-2 \pi i}{2})= - \sinh (z/2)$, we have $F(z- 2 \pi i)= F(z)$. Now, we know that
\begin{align}
    F(t)= \omega_0 (j(t)j(0))
\end{align}
We also know that 
\begin{align}
    F(t- 2 \pi i)&= - \frac{1}{8 \pi^2} \frac{1}{\sinh^2 (t/2)} \\
    &= - \frac{1}{8 \pi^2} \frac{1}{\sinh^2 (-t/2)} \\
    &=\omega_0 (j(-t) j(0) )\\
    &= \omega_0 (j(0) j(t))  \label{KMS verify}
\end{align}
(\ref{KMS verify}) is exactly the KMS condition with $\beta= 2 \pi$. If we restore the surface gravity $\kappa$, i.e, let $V= V_0 e^{\kappa v}$ and repeat the whole exercise, we will obtain that $\omega_0$ is a KMS state with $\beta = \frac{2 \pi}{\kappa}$. Furthermore, we can also work out the covariance in logarithmic coordinate $v$. W first define the function $C(s)$ as
\begin{align}
    C(s)= - \frac{1}{8 \pi^2} \frac{1}{\sinh^2\big( \frac{s-i \epsilon}{2} \big)}
\end{align}
Its Fourier decomposition is
\begin{align}
    C(s) = \frac{1}{2 \pi} \int^{\infty}_{- \infty} d \nu  S(\nu) e^{- i \nu s }
\end{align}
We want to find $S(\nu)$. Since the $U(1)$ chiral current has conformal weight one, we also have
\begin{align}
    [j (v), j(v')]= i \delta' (v-v') \label{ccr in v}
\end{align}
Consequently,  evaluating (\ref{ccr in v}) in vacuum gives
\begin{align}
    C(s)- C(-s)= i \delta'(s)= \frac{1}{2 \pi} \int^{\infty}_{- \infty} d \nu  \: \nu e^{-i \nu s}
\end{align}
So, it follows that
\begin{align}
    S(\nu)-S(- \nu)= \nu
\end{align}
KMS condition also implies that $S(- \nu)=e^{- 2 \pi \nu}S(\nu)$. Therefore, we obtain
\begin{align}
    S(\nu) = \frac{\nu}{1-e^{- 2 \pi \nu}}
\end{align}
and 
\begin{align}
    C(s)= \frac{1}{2 \pi } \int^{\infty}_{- \infty} d \nu  \frac{\nu}{1-e^{- 2 \pi \nu}} e^{-i \nu s}
\end{align}
Now, we also define the smeared field for $j(v)$ as
\begin{align}
    j_{v} (f) = \int^{\infty}_{- \infty} dv \: f(v) \: j (v)
\end{align}
In terms of smeared fields, the two point function is
\begin{align}
    \omega_{0} (j_{v} (f) j_{v} (g))= \frac{1}{2 \pi} \int^{\infty}_{- \infty} \frac{\nu}{1- e^{- 2 \pi \nu}}  \overline{\hat{f}(\nu)} \hat{g} (\nu)
\end{align}
Now, we can extract its real part to obtain the covariance (using $\hat{f}(-\nu)= \overline{\hat{f}(\nu)}$)
\begin{align}
    \mu (f,g)= \frac{1}{4 \pi} \int^{\infty}_{- \infty} d \nu \: \nu \coth(\pi \nu) \overline{\hat{f}(\nu)} \hat{g}(\nu) , \quad \omega\vert_{V>0} (W(f))=e^{- \mu (f,f)/2}  \label{key complicated cequation}
\end{align}

An important property of the chiral $U(1)$ current is that the modular automorphism group associated with the vacuum state $\omega_0$ restricted to the positive half-line algebra $\mathcal{M}{\geq 0}$ is geometrically implemented by the standard boost flow, as follows from the Bisognano--Wichmann property  \cite{Bisognano_1975,1993CMaPh.156..201B}. More precisely, the modular automorphisms act as dilations of the affine null coordinate $V$ (here we let $\kappa=1$),
\begin{align}
\sigma_s^{\omega_0}(J(V))
=e^{-2\pi s}J(e^{-2\pi s}V),\quad \sigma^{\omega_0}_s (W(f))=W(f_s), \: f_s= f(e^{2 \pi s} V)
\end{align}
In this simple model, we can also check the centralizer of the vaccum state on $\mathcal{M}_{\geq V_1}$ explicitly. We note that the calculation  uses from a standard result, namely the centralizer $\mathcal{C}_{\omega_0} (\mathcal{W}_{\geq V_1})$ is equal to the fixed point algbera under modular flow $\mathcal{C}_{fixed} (\mathcal{W}_{\geq V_1})$, i.e.,
\begin{align}
    \mathcal{C}_{\omega_0} (\mathcal{W}_{\geq V_1})= \mathcal{C}_{fixed} (\mathcal{W}_{\geq V_1}), \quad \mathcal{C}_{fixed} (\mathcal{W}_{\geq V_1})= \{ A \in \mathcal{W}_{\geq V_1}  : \: \sigma^{\omega_0}_s (A)=A \}
\end{align}
For further details, see section 4.2 of \cite{Sia:2025rux}. For later purpose, we want to show that the centralizer $\mathcal{C}_{\omega_0}(\mathcal{W}_{\geq V_1})$ is trivial. We note that for a cut at $V_c=0$, the modular automorphism group acts geometrically through $V \mapsto e^{- 2 \pi s} V$. For a cut at arbitrary affine endpoint $V_c$, the same modular flow acts instead by 
\begin{align}
    V \mapsto V_c + e^{- 2\pi s} (V-V_c)     \label{Vc flow}
\end{align}
To see why (\ref{Vc flow}) is true, we introduce a shifted coordinate
\begin{align}
    X= V- V_c
\end{align}
The half line $V > V_c$ becomes $X>0$. Since the vacuum is invariant under affine translation in $V$, its correlations in the $X$ coordinate have exactly the same form as for the original half-line. Moreover, modular flow dilates $X$, i.e., $X \mapsto e^{-2 \pi s} X$. Switching to $V$ coordinate, we obtain $V- V_c  \mapsto e^{-2 \pi s} (V-V_c)$, which is exactly (\ref{Vc flow}). We also recall that the original boost coordinate is
\begin{align}
    v= \log \big(\frac{V}{V_*} \big )
\end{align}
where $V_*$ is a constant. Under modular flow around $V_c$, we have
\begin{align}
    V \mapsto \log \big( \frac{V_c + e^{- 2 \pi s}(V-V_c)}{V_*}   \big)
\end{align}
So, if instead we define the cut dependent logarithmic coordinate as 
\begin{align}
    v_c = \log \big( \frac{V- V_c}{V_*}  \big)
\end{align}
then similarly we have $v_c \mapsto v_c - 2 \pi s$. Hence, each cut has its own natural modular time. Since the vacuum is invariant under affine translations, the modular dynamical systems associated with different cuts are unitarily equivalent. Consequently, if we can show that $\mathcal{C}_{\omega_0} (\mathcal{W}_{\geq 0})= \mathbb{C} \mathbb{1}$, then it automatically follows that $\mathcal{C}_{\omega_0} (\mathcal{W}_{\geq V_1})= \mathbb{C} \mathbb{1}$.

\par Now, \textbf{we want to show that  $\mathcal{C}_{\omega_0} (\mathcal{W}_{\geq 0})= \mathbb{C} \mathbb{1}$} . We recall that the one-paricle Hilbert space is $\mathfrak{h}= L^2 (\mathbb{R}_+, d \omega)$. A one-particle state is specified by a wavefunction $h(\omega), \omega>0$ with norm $\vert \vert h \vert \vert^2= \int^{\infty}_{0} d \omega \: \vert h \vert^2$. Note that a real test function $f(V)$ determines a one-particle wavefunction $h_f (\omega)= \sqrt{\frac{\omega}{2 \pi}} \hat{f}(\omega)$. A dilation acts on the one-particle wavefunction by 
\begin{align}
    (U_t h) (\omega)= e^{-t/2} h (e^{-t} \omega)  \label{dilation one particle h}
\end{align}
We can verify that $U_t$ is unitary.
\begin{align}
    \vert \vert U_t h \vert \vert^2 = \int^{\infty}_{0} d \omega \: \vert e^{-t/2} h (e^{-t} \omega) \vert^2= e^{-t} \int^{\infty}_{0} d \omega \vert h (e^{-t} \omega) \vert^2= \int^{\infty}_{0} d \omega' \vert h (\omega') \vert^2= \vert \vert h \vert \vert^2
\end{align}
In the third equality, we let $\omega'= e^{-t} \omega$. So, indeed $U_t$ is unitary and it preserves Hilbert space norm. The action of dilation (\ref{dilation one particle h}) is somewhat inconvenient because the group action acts by rescaling $\omega$. We can simplify it by introducing $x = \log \omega$ and we define a map $\mathfrak{U}: \mathfrak{h} \rightarrow L^2 (\mathbb{R}, dx)$ by
\begin{align}
    \mathfrak{U}h (x)= e^{x/2} \: h (e^x)
\end{align}
We can also check that $\mathfrak{U}$ is an unitary:
\begin{align}
    \vert \vert \mathfrak{U} h \vert \vert^2= \int^{\infty}_{- \infty} d \omega \: \omega^{-1} \omega \: \vert h (\omega ) \vert^2 = \vert \vert h \vert \vert^2
\end{align}
We calculate the transformed dilation action
\begin{align}
    (\mathfrak{U} U_th) (x)= e^{(x-t)/2} h (e^{x-t})= (\mathfrak{U}h) (x-t)
\end{align}
By letting $h (\omega)= \mathfrak{U}^{-1} (\psi (x)), \psi(x) \in L^2 (\mathbb{R}, dx)$, we obtain
\begin{align}
   ( \mathfrak{U} U_t \mathfrak{U}^{-1} \psi) (x)= \psi (x-t)
\end{align}
Our \textbf{ first small goal} in proving $\mathcal{C}_{\omega_0} (\mathcal{W}_{\geq 0})$ is to show that
\begin{align}
    \lim_{t \rightarrow 0}\langle h, U_t k \rangle \rightarrow 0, \quad h,k \in \mathfrak{h}   \label{first mixing result}
\end{align}
We denote $\psi = \mathfrak{U}h, \chi= \mathfrak{U}k$, then we have
\begin{align}
    \langle h, U_t k \rangle= \langle \psi, \mathfrak{U}U_t \mathfrak{U}^{-1} \chi \rangle= \int^{\infty}_{- \infty} dx \bar{\psi} (x) \chi (x-t)
\end{align}
We introduce the Fourier transform
\begin{align}
    \hat{\psi} (p) =\frac{1}{\sqrt{2 \pi}} \int_{- \infty}^{\infty} dx \: e^{-ip x} \psi(x)
\end{align}
Then, translations become multiplication by a phase
\begin{align}
    \widehat{\chi (\cdot -t)}(p)= e^{-i p t } \hat{\chi} (p)
\end{align}
By using Cauchy Schwarz inequality, one can also show that $F(p)= \overline{\hat{\psi}(p)} \hat{\chi} (p) \in L^1 (\mathbb{R})$. Thus, we can apply Riemann-Lebesgue lemma to obtain $\lim_{t \rightarrow \infty} \int^{\infty}_{- \infty} dp \: F(p) e^{-i pt}=0$. This is equivalent to (\ref{first mixing result}). So, we achieve our first small goal.

\par Next, the \textbf{second small goal} is to show \textbf{Weyl operator mixing} 
\begin{align}
    \lim_{t \rightarrow \infty} \omega_0 (A \alpha_t (B))= \omega_0 (A) \omega_0 (B), \quad A,B \in \mathcal{W}_{\geq 0}  \label{Weyl operator mixingxxx}
\end{align}
where $\alpha_t$ is the dilation automorphism. Under this dilation, we have
\begin{align}
    \alpha_t (W(g))= W(g_t), \quad h_{g_t}= U_t h_g
\end{align}
We calculate
\begin{align}
    \omega_0 \big(W(f) \alpha_t (W(g)) \big)&=\omega_0 \big(   W(f) W(g_t)\big) \\
    &= e^{- \frac{i}{2} \sigma (f, g_t)} \: e^{- \frac{1}{2} \vert \vert h_f + h_{g_t} \vert \vert^2}
\end{align}
We can expand the norm as $\vert \vert h_f + h_{g_t} \vert \vert^2 = \vert \vert h_f \vert \vert^2 + \vert \vert h_{g_t} \vert \vert^2 + 2 \Re{\langle h_f, h_{g_t}} \rangle$. Moreover, unitary gives $\vert \vert h_{g_t} \vert \vert^2 = \vert \vert h_g \vert \vert^2$ and we know that $\omega_0 (W(f))= e^{- \frac{1}{2} \vert \vert h_f \vert \vert^2}$. Therefore, we obtain
\begin{align}
    \omega_0 \big(W(f) \alpha_t (W(g)) \big) = \omega_0 (W(f)) \: \omega_0 (W(g))\: e^{- \Re{\langle h_f, U_t h_g \rangle}}e^{- \frac{i}{2} \sigma (f, g_t)}
\end{align}
Since the symplectic form is proportional to the imaginary part of the one-particle inner product, $\lim_{t \rightarrow \infty} \langle h_f, U_t h_g \rangle=0$ implies both $\lim_{t \rightarrow \infty}\Re{\langle h_f , U_t h_g \rangle}=0$ and $\lim_{t \rightarrow \infty}\sigma (f, g_t)=0$. So, we obtain (\ref{Weyl operator mixingxxx}). Since the collection of Weyl operator is strongly dense in $\mathcal{W}_{\geq 0}$ and finite Weyl combination can approximate vectors of the form $A \ket{0}$ with $A \in \mathcal{W}_{\geq 0}$.  So, (\ref{Weyl operator mixingxxx}) can be extended to entire $\mathcal{W}_{\geq 0}$.

\par Now, we can \textbf{start to show $\mathcal{C}_{\omega_0}(\mathcal{W}_{\geq 0}) = \mathbb{C} \mathbb{1}$}. Equivalently, we want to show that $\mathcal{C}_{fixed} (\mathcal{W}_{\geq 0})=\mathbb{C} \mathbb{1} $. Suppose that $X \in \mathcal{C}_{fixed} (\mathcal{W}_{\geq 0}) $, then $\sigma^{\omega_0}_s (X)=X$. We consider
\begin{align}
    \omega_0 (X^* \sigma^{\omega_0}_s (X))= \omega_0 (X^* X)
\end{align}
but (\ref{Weyl operator mixingxxx}) tells that 
\begin{align}
    \lim_{s \rightarrow \infty} \omega_0 (X^* \sigma_s (X))= \omega_0 (X^*) \omega_0 (X)= \vert \omega_0 (X) \vert^2
\end{align}
So, we obtain
\begin{align}
    \omega_0 (X^* X)= \vert \omega_0 (X) \vert^2  \label{final mixing}
\end{align}
Now, we let $Y= X- \omega_0 (X) \mathbb{1}$. (\ref{final mixing}) will imply that $\omega_0 (Y^* Y)=0$. Since vacuum is faithful, it also implies that $Y=0$. This is equivalent to $X= \omega_0 (X) \mathbb{1}$. So, every element of the centralizer is therefore a scalar
\begin{align}
    \mathcal{C}_{\omega_0} (\mathcal{W}_{\geq 0}) = \mathbb{C} \mathbb{1}
\end{align}
This also implies that
\begin{align}
    \mathcal{C}_{\omega_0} (\mathcal{W}_{\geq V})=\mathbb{C} \mathbb{1}, \quad \forall V >0
\end{align}
Finally, it is also instructive to discuss a nice property. Suppose that $B \in \mathcal{W}_{\geq 0}$ is a bounded operator which is an eigenoperator of the modular automorphism group with eigenvalue
\begin{align}
    \sigma^{\omega_0}_s (B)= e^{i \nu s} B, \quad \nu \in \mathbb{R}
\end{align}
We assume that $\nu \neq 0$. Since vacuum state is invariant under the modular flow, we have
\begin{align}
    \omega_0 (\sigma_s^{\omega_0} (B))= \omega_0 (B)
\end{align}
but eigenprojector relation gives
\begin{align}
    \omega_0 (\sigma_s^{\omega_0} (B))= e^{i \nu s} \omega_0 (B)
\end{align}
This implies that $(e^{i \nu s} -1) \omega_0 (B)=0$. By our assumption that $\nu \neq0$, we know that $\omega_0 (B)=0$. Applying (\ref{Weyl operator mixingxxx}), we have
\begin{align}
    \lim_{s \rightarrow \infty} e^{i \nu s}\omega_0 (B^* B)=\lim_{s \rightarrow \infty} \omega_0 (B^* \sigma^{\omega_0}_s (B))=\omega_0 (B^*) \omega_0 (B)=0 \label{star}
\end{align}
Since the phase has modulus one,  we have $\vert e^{i \nu s} \omega_0 (B^* B) \vert = \omega_0 (B^* B) $. So, (\ref{star}) is only possible if $\omega_0 (B^* B)=0$. By faithfulness of vacuum, we obtain $B=0$. This means that there is \textbf{no nonzero frequency eigenoperators}. 

\subsubsection{Setup and Interaction}
Now, we want to describe the  setup for our adapted FP model. As we have discussed in detail, the Weyl algebras $\mathcal{W}_{\geq V}$ generated by $U(1)$ chiral current algebra play the role of horiozn cut algebra $\mathcal{M}_{\geq \lambda}$.  These algebras are naturally parametrized by the affine null coordinate $V$ along the future horizon. However, for our purposes, it is more convenient to work with the time parameter $v= \log \big( \frac{V}{V_*} \big)$ adapted to the dilation symmetry of the horizon. We recall that $j (v)= VJ(V)$. We can smear $j(v)$ with a test function $f$
\begin{align}
    j_v(f) &= \int_{- \infty}^{\infty} dv \: j(v) f(v) \\
    &= \int^{\infty}_{0}dV\:  J(V) f \bigg( \log \big(\frac{V}{V_*} \big) \bigg) \\
    &= \int^{\infty}_0 dV J(V) F(V) = J(F)    \label{identification good}
\end{align}
where we define $F(V)= f \bigg( \log \big(\frac{V}{V_*} \big) \bigg)$. So, we see that $j_v (f)= J(F)$. We can also define a Weyl algebra $\mathcal{Y}$ 
\begin{align}
    \mathcal{Y}= \bigg\{ W_v (f):= e^{i j_v (f)}   \vert \: f\in C^{\infty}_{c} (\mathbb{R})   \bigg \}''
\end{align}
Due to (\ref{identification good}), we know $W_v (f)= W(F); \: F=f \bigg( \log \big(\frac{V}{V_*} \big) \bigg)$. Therefore we know that $\mathcal{Y}= \mathcal{W}$ and we also have
\begin{align}
    \mathcal{Y}_{\geq v}= \mathcal{W}_{\geq V_* e^v}
\end{align}
Then, PDP also follows immediately
\begin{align}
    \mathcal{Y}_{\geq v_2} \subsetneq \mathcal{Y}_{\geq v_1}, \quad v_2 > v_1
\end{align}
and all properties we discussed before are inherited to $\mathcal{Y}_{\geq v}$. In particular, $\mathcal{Y}_{\geq v}= \mathcal{W}_{\geq V_* e^v}$ also has trivial centralizer with respect to vacuum and no nonzero frequency eigenoperators.

On the other hand, for the system undergoing decoherence, the `` atomic system'' in the FP model naturally plays the role of the charged particle
in the DSW setting.  Accordingly, the finite-dimensional Hilbert
space $\mathcal{H}_A$ is identified with the
effective Hilbert space of the charged particle,
while the atomic density matrix $\Omega$ represents
its quantum state. For simplicity, we restrict attention to the
two-dimensional subspace spanned by the orthonormal
states $\{|\psi_1\rangle,|\psi_2\rangle\}$ introduced
earlier, which represent the two alternative
branches of the charged-particle configuration.
We therefore take
\begin{align}
    \mathcal{H}_A
    &=
    \operatorname{span}
    \{|\psi_1\rangle,|\psi_2\rangle\},
    \qquad
    \dim\mathcal{H}_A=M=2.
\end{align}
The corresponding family of mutually orthogonal
projections $\{Q_n\}_{n=1}^{2}$, which forms a
resolution of the identity on $\mathcal{H}_A$,
is naturally chosen to be
\begin{align}
    Q_1=|\psi_1\rangle\langle\psi_1|,
    \qquad
    Q_2=|\psi_2\rangle\langle\psi_2|,
    \qquad
    Q_1+Q_2=\mathbb{1}_{\mathcal{H}_A}.
\end{align}
Now, for the \textbf{interaction}, we imitate the method in the original discrete FP model by introducing an unitary operator: we let $F_{m, T} \in C^{\infty}_{0}((0, T))$ be prescribed field smearings for the two charged particle branches and set
\begin{align}
    \mathrm{U}_{T}= \sum_{m=1}^{2}Q_m \otimes W(F_{m,T})
\end{align}
As contrast to the original FP model, we do not divide the duration into many collison steps and we just do it for one shot. We let the initial step $\omega_{in}= \omega_A \otimes \omega_0$ (where $\omega_A (\cdot_\mathcal{A}) =\Tr (\Omega \cdot _{\mathcal{A}} )$) and our initial algebra is $\mathcal{E}_{\geq 0}= \mathcal{A} \otimes \mathcal{Y}_{\geq 0}$. Suppose after a sufficiently long dilation time $T$, our experiment has been completely finished, then we let our final algebra to be $\mathcal{E}_{\geq T} \simeq \mathcal{A} \otimes \mathcal{Y}_{T }$ ($\simeq$ here means unitarily equivalent) and we introduce a $*$ homomorphism $\mathfrak{s}: \mathcal{E}_{\geq 0} \rightarrow \mathcal{E}_{\geq T}$ as
\begin{align}
    \mathfrak{s} (A \otimes B)= \mathrm{U}_{T} (A \otimes \alpha_T (B)) U^{*}_T, \quad B \in \mathcal{Y}_{\geq 0}
\end{align}
Here, $\alpha_T$ is the dilation automorphism. Then, we define the out state $\omega_{out}= \omega_{in} \circ \mathfrak{s}$, i.e., we denote $\mathfrak{s}(A \otimes B)=E \in \mathcal{E}_{ \geq T}$ and we have
\begin{align}
    \omega_{out} (E)=( \omega_A \otimes \omega_0) (\mathrm{U}_T (A \otimes \alpha_T (B)) \mathrm{U}^{*}_{T})
\end{align}
Direct evaluation gives
\begin{align}
    \mathrm{U}_T (A \otimes \alpha_T (B)) \mathrm{U}^{*}_{T}= \sum_{m,n=1}^{2} \omega_A (Q_m A Q_n) \: \mathfrak{S}^T_{mn} (B)
\end{align}
where 
\begin{align}
    \mathfrak{S}^{T}_{mn} (B)= \omega_0 \bigg( W(F_{m,T})\alpha_T (B) \: W (F_{n,T})^*   \bigg)
\end{align}
Now, we let $B= W(h)$ with $h \in C^{\infty}_{0} ((0, \infty))$. We know that $\alpha_T (h)= h_T$. In the dilation/boost time representation, we have $h_T(v)= h(v-T)$ and $supp(h_T) \subset [T, \infty)$. So, we need to calculate
\begin{align}
    \mathfrak{S}^T_{mn} (W(h))= \omega_0 \bigg( W(F_{m,T}) W(h_T)  W(F_{n,T})^*   \bigg)  \label{things we evaluate}
\end{align}

Since $supp (h_T)$ and $supp(F_{m,T})$ are disjoint, we know that $\sigma(F_{m,T}, h_T)=0$. So, we have
\begin{align}
    W(F_{m,T}) W(h_T) W(F_{n,T})^*= e^{\frac{i}{2} \sigma (F_{m,T   }, F_{n,T})} \: W( h_T + \Delta F_{mn,T}), \quad \Delta F_{mn,T}= F_{m,T}- F_{n,T}    \label{three W}
\end{align}

Evaluating (\ref{three W}) in vacuum state, we obtain
\begin{align}
    \mathfrak{S}^T_{mn} (W(h))= e^{\frac{i}{2} \sigma (F_{m,T}, F_{n,T})} e^{- \frac{1}{2} \mu (\Delta F_{mn,T}, \Delta F_{mn},T)} \: e^{- \frac{1}{2} \mu (h_T, h_T)} \: e^{- \mu (h_T, \Delta F_{mn,T})}
\end{align}
Since the vaccum state is invariance under boost/dilation, we have $\mu(h_T, h_T)= \mu (h,h)$. This implies that
\begin{align}
    e^{- \frac{1}{2} \mu (h_T, h_T)}= \omega_0 (W(h))
\end{align}
We also note that $e^{\frac{i}{2} \sigma (F_{m,T}, F_{n,T})} e^{- \frac{1}{2} \mu (\Delta F_{mn,T}, \Delta F_{mn},T)}= \omega_0 (W(F_{m,T}) W(F_{n,T})^*)$. We define the Gram matrix $g_{mn} (T)$ as
\begin{align}
    g_{mn} (T)= \omega_0 (W(F_{m,T})W(F_{n,T})^*)
\end{align}
Therefore, we obtain
\begin{align}
    \mathfrak{S}^T_{mn} (W(h))= g_{mn} (T) \: \omega_{0} \big(W(h) \big) \: e^{- \mu(h_T, \Delta F_{mn, T})}
\end{align}
This is analogous to the original discrete FP model where the Gram matrix $g_{mn} (T)$ is absorbed into the ``atomic'' system. However, for our adapted model, we see that there is an extra cross term  $e^{- \mu(h_T, \Delta F_{mn,T})}$. As we will argue later, this cross term has negligible effect at large $T$ limit.

\par Now, we first analyze the Gram matrix $g_{mn} (T)$. For $m=n$, we have $W(F_{n,T}) W(F_{n,T}) ^*\sim W(0)=1$, so we know $g_{nn}=1$.  To analyze the off diagonal term $g_{12}$, we specify the symmetric two-branch interaction:
\begin{align}
    F_{1,T} (v)= \frac{\gamma}{2} s_T (v), \quad F_{2,T} (v)= -\frac{\gamma}{2} s_T (v) 
\end{align}
where $\gamma$ is a constant and $s_T (v)$ is a pulse for which its width $\sim T$. Their difference $\Delta F_{12,T}= - \gamma s_T $.  So, we have $\mu (\Delta F_{12}, \Delta F_{12})= \gamma^2 \mu (s_T, s_T)$. Furthermore, since $\sigma (F_{1,T}, F_{2,T}) \sim \sigma (s_T, s_T)=0$, the Gram coefficient is real and positive
\begin{align}
    g_{12} (T)= e^{- D_T}, \quad D_T= \frac{1}{2} \mu \bigg(   \Delta F_{12, T}, \Delta F_{12, T} \bigg)
\end{align}
Using (\ref{key complicated cequation}), we know that
\begin{align}
    D_T = \frac{\gamma^2}{8 \pi} \int^{\infty}_{- \infty} d \nu \: \nu \coth (\pi \nu) \: \vert \hat{s}_T(\nu) \vert^2
\end{align}
Similarly as what we have done in the previous section, to understand the leading  behaviour, we just need to consider a sharp rectangular profile
\begin{align}
    s_T (v)= \chi_{[0, T]} (v)
\end{align}
Its Fourier transform is 
\begin{align}
    \hat{\chi}_{[0,T]} (\nu)&= \int_{0}^{T} dv \: e^{i \nu v}= \frac{e^{i \nu T}-1}{i \nu}= 2 e^{i \frac{\nu T}{2}} \frac{\sin \big( \frac{\nu T}{2} \big)}{\nu}
\end{align}
Taking its modulus, we obtain
\begin{align}
    \vert \hat{\chi}_{[0,T]} (\nu) \vert^2= \frac{4 \sin^2 \big(  \frac{\nu T}{2} \big)}{\nu^2} = \frac{2(1-\cos \nu T)}{\nu^2}
    \label{modulus square pulse}
\end{align}
Substituting (\ref{modulus square pulse}) into the covariance, we obtain
\begin{align}
    D_T &= \frac{\gamma^2}{4 \pi} \int^{\infty}_{- \infty} d \nu \:\nu \coth (\pi \nu)  \frac{1- \cos \nu T}{\nu^2} \\
    &= \frac{\gamma^2}{2 \pi} \int^{\infty}_{0} d \nu \; \nu \coth (\pi \nu) \frac{1-\cos \nu T}{\nu^2}
\end{align}
In the second equality, we use that the integrand is even. Now, we divide frequency scales into $3$ regions, namely $(0, \nu_{IR}), (\nu_{IR}, \nu_{UV})$ and $(\nu_{UV}, \infty)$. We will analyze each case separately.

\par In the \textbf{deep infrared region }$0< \nu \ll 1$, we have $\coth \nu \sim \frac{1}{\pi \nu}$ and 
\begin{align}
    D^{IR}_{T} \approx \frac{\gamma^2}{2 \pi^2} \int^{\nu_{IR}}_{0}  \frac{1- \cos \nu T}{\nu^2} \: d \nu
\end{align}
Letting $x= \nu T$, we obtain
\begin{align}
     D^{IR}_{T} \approx \frac{\gamma^2}{2 \pi^2} T\int^{\nu_{IR} T}_{0} \frac{1- \cos x}{x^2} d x \approx \frac{\gamma^2}{2 \pi^2} T\int^{\infty}_{0} \frac{1- \cos x}{x^2} d x= \frac{\gamma^2}{4\pi}T 
\end{align}
In the second approximation, we take $T \rightarrow \infty$ since we want to capture large $T$ behaviour. So, in the deep infrared region, we obtain linear growth in $T$
\begin{align}
    D^{IR}_{T} = \frac{\gamma^2}{4\pi} T + \mathcal{O}(1)
\end{align}

\par In the \textbf{intermediate region} $\nu_{IR}< \nu < \nu_{UV}$, we have
\begin{align}
    D_T^{mid} = \frac{\gamma^2}{2 \pi} \int^{\nu_{UV}}_{\nu_{IR}} d \nu  \frac{\coth (\pi \nu)}{\nu}   -\:  \frac{\gamma^2}{2 \pi} \int^{\nu_{UV}}_{\nu_{IR}} d \nu \: \frac{\coth (\pi \nu)}{\nu} \cos (\nu T)
\end{align}
For the first term,  since the integration interval is bounded away from both zero and infinity, the integral is finite and yield $\mathcal{O}(1)$ result independent of $T$. For the second term, we can apply Riemann-Lebesgue lemma to capture large $T$ behaviour
\begin{align}
    \lim_{T \rightarrow \infty}\int^{\nu_{UV}}_{\nu_{IR}} d \nu \: \frac{\coth (\pi \nu)}{\nu} \cos (\nu T) \rightarrow 0
\end{align}
Hence, in the intermediate region, we have $\mathcal{O}(1)$ output
\begin{align}
    D^{mid}_{T} =\mathcal{O} (1), \quad T \rightarrow \infty
\end{align}

In the \textbf{deep ultraviolet region}, we have $\coth (\pi \nu) \sim1$. So,
\begin{align}
    D^{UV}_{T}\approx \frac{\gamma^2}{2 \pi} \int_{\nu_{UV}}^{\infty} d \nu \:  \frac{1- \cos \nu T}{\nu}
\end{align}
This integral is logarithmic divergent. To see it explicitly, we introduce a $UV$ cutoff $\Lambda$
\begin{align}
    D^{UV}_{T}\approx  \frac{\gamma^2}{2 \pi}\int^{\Lambda}_{\nu_{UV}} d \nu \frac{1- \cos \nu T}{\nu} \sim \frac{\gamma^2}{2 \pi} \log (\frac{\Lambda}{\nu_{UV}}) + \mathcal{O} (1)
\end{align}
This logarithmic divergent is entirely artificial due to our use of sharp pulse which is discontinuous at $v=0$ and $v=T$.
Its Fourier transform decays only as $\hat{\chi}(\nu) \sim \frac{1}{\nu}$ in its oscillatory envelope. Meanwhile, the chiral-current spectral density $\nu \coth (\pi \nu)$ grows as $\sim \nu$. The corresponding product $\nu \coth (\pi \nu) \vert \hat{\chi}(\nu) \vert^2$ therefore behaves as
\begin{align}
    \nu \coth (\pi \nu) \vert \hat{\chi}(\nu) \vert^2 \sim \frac{\textit{oscillatory factor}}{\nu}
\end{align}
The integral of this high-frequency tail diverges logarithmically. A physically natural way to remove the divergence is to have a smooth pulse, i.e, allow the interaction to switch on and off over a finite time  $\delta >0$. For example, we can introduce a smooth normalized switching kernel $r_{\delta} (v)$ with characteristic width $\delta$ and define
\begin{align}
    s_{T} (v)= \int^{T}_{0} r_\delta  (v -u) du
\end{align}
Its Fourier transform is 
\begin{align}
    \hat{s}_T (\nu)= \hat{r}_{\delta} (\nu) \frac{1- e^{-i \nu T}}{i \nu}
\end{align}
Therefore, we have
\begin{align}
    D_T = \frac{\gamma^2}{2 \pi} \int^{\infty}_{0} d \nu \: \frac{\nu \coth(\pi \nu)}{\nu^2} \vert\hat{r}_{\delta}(\nu) \vert^2 (1- \cos \nu T)
\end{align}
In particular, $\hat{r}_{\delta}(\nu)$ is smooth and it decays rapidly at large frequency. The ultraviolet integral is now convergent. At low frequencies, we have $\hat{r}_{\delta}(0)=1$, so the linear $T$ contribution in the deep IR region remains unchanged.

Hence, for every fixed finite switching duration $\delta$, we obtain
\begin{align}
   D^{(\delta)}_{T}= \frac{\gamma^2}{4\pi} T + \mathcal{O}(1)
\end{align}
and thus
\begin{align}
    g_{12} (T)= e^{- \frac{\gamma^2}{4\pi}T + \mathcal{O}(1)} 
\end{align}
Now, we want to analyze the large $T$ behaviour of the cross term 
\begin{align}
    K_{mn}^{h_T} (T):= e^{- \mu (h_T, \Delta F_{mn, T})}  \label{aaaaaaa}
\end{align}
 We first note that for $m=n$, $\Delta F_{nn}=0$ and so $K_{nn}^{h_T} (T)=1$. So, the non-trivial part  is still the off-diagonal part $K_{12}^{h_T} (T)$ (note that $K_{12}^{h_T}(T)=(K_{21}^{h_T} (T))^{-1}$ because $\Delta F_{12}=- \Delta F_{21}$). Since $supp (h_T)$ and $supp (\Delta F_{mn,T})$ is disjoint, $\sigma (h_T, \Delta F_{mn, T})=0$ and
 \begin{align}
     \mu (h_T, \Delta F_{mn,T})&= - \frac{1}{8 \pi}\int_{T}^{\infty} dv' \: \int_{0}^{T} dv  \: h_T (v')  \Delta F_{mn, T} (v) \frac{1}{ \sinh^2 \big( \frac{v-v'}{2}\big)} \\
     &=- \frac{\gamma}{8 \pi } \int_{T}^{\infty} dv' \: \int_{0}^{T} dv \:  \frac{h(v'-T)  \: s_T (v)}{ \sinh^2 \big( \frac{v-v'}{2}\big)}
\end{align}
We can introduce coordinates $u = T-v$ and $w=v'-T$. In term of these coordinates, we have
\begin{align}
    \mu (h_T, \Delta F_{12,T})= - \frac{\gamma}{8 \pi} \int^{\infty}_{0} d \omega \: h(\omega) \: \int^{T}_{0} d u \: \frac{s_T (T-u)}{\sinh^2 \bigg( \frac{(u+w)}{2}  \bigg)}
\end{align}
The interpretation here is that $u$ measures how far the branch disturbance lies in the past of the final cut $T$ while $w$ measures how far the future observable lies beyond that cut. To capture the leading large $T$ behaviour, we again let $s_T$ to be sharp rectangular profile. We introduce the step function
\begin{align}
    q(u) =
\begin{cases}
  0 & \text{if } u \leq 0, \\
  1  & \text{if } u > 0.
\end{cases}
\end{align}
 We can let $s_T (v)= q(v)q(T-v)$.  So, $s_T (T-u)= q(T-u) q(u)$. Therefore, we obtain
 \begin{align}
      \mu (h_T, \Delta F_{12,T})=- \frac{\gamma}{8 \pi} \int^{\infty}_{0} d \omega \: h(\omega) \frac{q(T-u) q(u)}{\sinh^2 \bigg(  \frac{u+w}{2} \bigg)}
 \end{align}
 In large $T$ limit, we can let $q(T-u) \rightarrow1$. So, at \textbf{large $T$}, we have
 \begin{align}
     \mu (h_T, \Delta F_{12,T}) \approx - \frac{\gamma}{8 \pi} \int^{\infty}_{0} d w \: h(w) \: H_q (w)
 \end{align}
 where the switching dependent function $H_q (w)$ is
 \begin{align}
     H_q (w)= \int^{\infty}_{0} du \: \frac{q(u)}{\sinh^2 \big(  \frac{u+w}{2} \big)}
 \end{align}
 This function can also be evaluated analytically (use $\frac{d}{du} \coth (\frac{u+w}{2})=- \frac{1}{2 \sinh^2 ((u+w)/2)}$)
 \begin{align}
     H_q (w)= \frac{4}{e^w -1}
 \end{align}
 Therefore, at \textbf{large $T$}, we obtain
 \begin{align}
     \mu (\Delta F_{12,T}, h_T) \approx - \frac{\gamma}{2 \pi} \int^{\infty}_{0} \frac{h(\omega)}{e^{\omega}-1} d \omega \sim \mathcal{O}(1)
 \end{align}
 We know it is of order $1$ because $h(\omega)$ is compactly supported. Now, we can compare the cross term $K_{mn} (T)$ with the Gram matrix $g_{mn} (T)$ in large $T$ limit by comparing their exponent
 \begin{align}
    \lim_{T \rightarrow \infty}  \frac{\mu (\Delta F_{12,T}, h_T)}{\log g_{12} (T)} \sim \frac{\mathcal{O}(1)}{\mathcal{O} (T)} \rightarrow0
 \end{align}
 Hence, at large $T$, the cross term $K_{mn}^{h_T} (T) $ can be safely ignored and we have
 \begin{align}
     \mathfrak{S}^{T}_{mn} (W(h)) \approx g_{mn} (T) \omega_0 (W(h)) \quad \textit{at large }T 
 \end{align}
 Therefore, at large holding time $T$, we can absorb the Gram matrix $g_{mn} (T)$ into the charged particle's density matrix to obtain (here we ignore the free evolution $V$ of the `` atomic'' system in the original FP model because it plays no role in DSW setting)
 \begin{align}
     \Omega (T) \approx \sum_{n,m=1}^{2} g^{nm} (T) Q_n \Omega_{\textit{in}} Q_m \label{large T beh of Omega}
 \end{align}
 where the initial density matrix is a pure state 
 \begin{align}
     \Omega_{\mathrm{in}} = \ket{\Psi}\bra{\Psi}= \frac{1}{2} \begin{pmatrix}
         1 & 1\\ 1 &1
     \end{pmatrix}
 \end{align}
 In matrix notation, (\ref{large T beh of Omega}) is
 \begin{align}
     \Omega (T) \approx \frac{1}{2} \begin{pmatrix}
         1 &e^{- \Gamma T } \\
        e^{- \Gamma T  } &1
     \end{pmatrix}, \: \Gamma= \frac{\gamma^2}{ 4\pi}, \quad \textit{at large}\: T  \label{large T matix}
 \end{align}
 We therefore recover the characteristic
exponential decoherence behavior found in the
DSW framework.
  We note that (\ref{large T matix}) is only a good approximation for large holding time $T$. It is also obvious that (\ref{large T matix}) is a solution to the deterministic equation

 \begin{align}
    \frac{\partial \Omega}{\partial T}= \frac{\Gamma}{2} \bigg( \sigma_z \Omega \sigma_z - \Omega   \bigg), \quad \sigma_z= \begin{pmatrix}
        1 & 0 \\
        0 & -1   \label{deterministic Lindbladddd eqn}
    \end{pmatrix}
\end{align}
In matrix notation, $Q_1 = \ket{\psi_1} \bra{\psi_1}$ and $Q_2 = \ket{\psi_2} \bra{\psi_2}$ can be written as
\begin{align}
    Q_1 = \begin{pmatrix}
        1 & 0 \\
        0 & 0
    \end{pmatrix}, \quad Q_2 = \begin{pmatrix}
        0 & 0 \\
        0 & 1
    \end{pmatrix}, \quad \sigma_z = Q_1-Q_2= 2Q_1- \mathbb{1}= \mathbb{1}-2Q_2
\end{align}
In terms of $Q_1$ and $Q_2$, (\ref{deterministic Lindbladddd eqn}) can be further expressed as
\begin{align}
    \frac{\partial \Omega}{\partial T} &= - \Gamma (Q_1 \Omega Q_2 + Q_2 \Omega Q_1) \\
    &= \Gamma \big( - (Q_1 + Q_2) \Omega (Q_1 + Q_2)+ Q_1 \Omega Q_1 + Q_2 \Omega Q_2 \big) \\
    &= \Gamma \sum_{n=1}^2 \bigg( Q_n \Omega Q_n -\frac{1}{2} \{ Q_n, \Omega \}  \bigg) 
\end{align}
To go from the second line to the third line, we use $Q_1 + Q_2 = \mathbb{1}$. Therefore, we see that the asymptotic decoherence law is
consistent with an effective GKSL description
of pure dephasing. This is analogous to the 
weak-coupling limit of the original FP model. However, it does not 
imply that the actual effective dynamics underlying
horizon-induced decoherence admits an equally
simple Markovian description. To describe the actual effective dynamics, one needs to take into account the effect of the cross term (\ref{aaaaaaa}). Including
this contribution, at time $T$, the global state without taking into account collapse postulate/events occurring yet  is
\begin{align}
    \sum_{m,n=1}^{2} \omega_A (Q_m \cdot_{\mathcal{A}} Q_n) \: g_{mn} (T) \; K_{mn}^{\circ _{\mathcal{Y}_{\geq T}}} (T)  \: \omega_0 (\cdot_{\mathcal{Y}_{\geq T}})  \label{effective state no collapse}
\end{align}
Crucially, the term $K_{mn}^{\circ_{\mathcal{Y}_{\geq T}}} (T)$ depends
non-trivially on the observable of the future horizon
algebra $\mathcal{Y}_{\geq T}$. Consequently, the joint state (\ref{effective state no collapse}) cannot, in general,
be expressed as a tensor product of the state of the
system undergoing decoherence and the vacuum state
on the future horizon algebra Indeed, the additional
cross term prevents the Gram coefficients $g_{mn} (T)$ from being
completely absorbed into the charged particle system  while
leaving an independent vacuum state on the future horizon.
It is in this sense that the usual Markovian description
in terms of a GKSL equation does not fully capture the
underlying effective dynamics even if we neglect all order $1$ corrections in Gram matrix $g_{mn} (T)$.
This failure of factorization is a consequence
of the nontrivial vacuum correlations between
the interaction region and the future horizon.

\par We are therefore led to conjecture that the phenomenon
observed in our adapted FP model reflects a more general
feature of the effective dynamics underlying horizon-induced
decoherence. In particular, our observation suggests
the presence of non-Markovian features in the underlying effective
dynamics, despite the apparently Markovian behavior of
the reduced system in the large holding-time limit.
\subsection{Implications}
The most profound implication of the original discrete FP model
is that it brings to light the intrinsic stochasticity of
quantum dynamics, thereby opening up the possibility of a
unified dynamical framework in which the occurrence of
quantum events and measurements are naturally incorporated.
In the previous subsection, we discussed the effective
dynamics of our adapted FP model intended to describe
horizon-induced decoherence. We now briefly comment on
the ontology of the dynamics, based on the criteria for
the occurrence of quantum events in the ETH approach.

\par
In the ETH approach to quantum mechanics, the structure
of actual events is determined by the center of the
centralizer of the physical state on the relevant
time-band algebra. For our adapted FP model, suppose
that, for $v<v_0$, the interaction is switched off
and the charged particle is prepared in the pure state
$\Omega_{\mathrm{in}}=|\Psi\rangle\langle\Psi|$,
so that
\begin{align}
    \Omega(v<v_0)
    =
    \Omega_{\mathrm{in}}
    =
    \frac{1}{2}
    \begin{pmatrix}
        1 & 1\\
        1 & 1
    \end{pmatrix}.
\end{align}

Suppose that the interaction is switched on at $v=v_0$.
As we evolve from $v_0$ to the next infinitesimal
time step $v_0+dv$, the relevant interacting algebra
is $\mathcal E_{\geq v_0+dv}
\simeq \mathcal A {\otimes}
\mathcal Y_{\geq v_0+dv}$, given by
\begin{align}
    \mathcal E_{\geq v_0+dv}
    &=
    U_{dv}
    \left(
        \mathcal A{\otimes}
        \mathcal Y_{\geq v_0+dv}
    \right)
    U_{dv}^{*}
    \nonumber\\
    &=
    \sum_{m,n=1}^{2}
    Q_m\mathcal A Q_n
    \otimes
    W(F_{m,dv})
    \mathcal Y_{\geq v_0+dv}
    W(F_{n,dv})^*.
\end{align}

The algebra $\mathcal E_{\geq v_0+dv}$ is equipped
with the state $\omega_A\otimes\omega_0$,
restricted to this algebra, where $\omega_A$
is associated with the initial density matrix
$\Omega_{\mathrm{in}}$. This description is
unitarily equivalent to the product algebra
$\mathcal A{\otimes}
\mathcal Y_{\geq v_0+dv}$ equipped with the
state $(\omega_A\otimes\omega_0)\circ\beta$,
where $\beta$ is defined by
\begin{align}
    \beta(A\otimes B)
    =
    U_{dv}(A\otimes B)U_{dv}^{*},
    \qquad
    A\in\mathcal A,\quad
    B\in\mathcal Y_{\geq v_0+dv}.
\end{align}

Therefore, to determine whether the first actual
event occurs after the interaction is switched on,
we need to find the center of the centralizer $ Z_{\omega_A\otimes\omega_0}
    \left(
        \mathcal E_{\geq v_0+dv}
    \right)$
of the state $\omega_A\otimes\omega_0$ on the
interacting algebra $\mathcal E_{\geq v_0+dv}$.
This is clearly not a straightforward/ easy task,
although it is not necessarily impossible.
We also refrain from carrying out this analysis
in the present paper. Moreover, even if we succeeded in identifying
the first actual event, analyzing the subsequent
events would become considerably more difficult,
since the physical state changes after each event
and the corresponding center of the centralizer
must be determined anew. At present, we do not
have sufficient control over this recursive process. Thus, even in this relatively simple model, the
axiomatic formulation of the ETH approach does
not, by itself, provide an explicit construction
of the full stochastic dynamics of quantum events.
A more complete understanding of this problem
is left for future investigation. Our  expectation is that the event structure of this adapted FP model should be rather rich, for the following reason.  We have shown that, when the horizon algebra is considered in isolation and equipped with the vacuum state, its centralizer is trivial, i.e, $\mathcal{C}_{\omega_0} (\mathcal{Y}_{\geq v})= \mathbb{C} \mathbb{1}, \: \forall v $. This is natural because the vacuum state $\omega_0$ is a KMS state with respect to the dilation/ boost time. In thermal equilibrium, no events should occur because  the system equilibrates. However, introducing an auxiliary system and couple it together with the thermal equilibrium system will disturb the equilibrium and thus lead to the occurence of events.

\par
Finally, it is also instructive to compare the occurrence
of quantum events in our adapted FP model with that in the
original discrete FP model. Recall that, in the original
FP model, the event structure is fully under control,
giving rise to a self-contained stochastic dynamics.
Let us briefly recall how this works. Suppose that, at discrete time $\tau$, the state on the
algebra $\mathcal{E}_{\geq \tau}$ is given by $\omega_{\tau}
    =
    \Tr(\Omega_{\tau}\,\cdot_{\mathcal A})
    \otimes \phi_{\vec{0}}$.
As we evolve from $\tau$ to $\tau+1$, the algebra
$\mathcal{E}_{\geq \tau+1}$ is given by
(\ref{explicit E algebrasss}), equipped with the
restricted state
$\omega_{\tau}|_{\mathcal{E}_{\geq \tau+1}}$. Now, owing to the factorization property
(\ref{key discovery for discrete FP}), the algebra
$\mathcal{E}_{\geq \tau+1}$ equipped with this
restricted state is \textbf{physically equivalent}  to the
description using the product algebra
$\mathcal{A}\otimes\mathcal{B}_{\geq \tau+1}$,
equipped with the state
\begin{align}
    \Tr\left(
        \Psi_{\tau}(\Omega_{\tau})
        \,\cdot_{\mathcal A}
    \right)
    \otimes \phi_{\vec{0}},
\end{align}
where $\Psi_{\tau}$ is the CPTP map defined in
(\ref{cptppp}). Consequently, the event structure can be determined
from the spectral projections of the updated atomic
density matrix $\Psi_{\tau}(\Omega_{\tau})$.
More precisely, let $  P_{\vec{0}}
    =
    |\Phi_{\vec{0}}\rangle
    \langle\Phi_{\vec{0}}|$
denote the support projection of the pure reference
state of the $R$-fields. On the support of this state,
the center of the centralizer is  given by 
\begin{align}
    Z_{\Psi_{\tau},\phi_{\vec{0}}}^{\mathrm{supp}}
    &=
    \left\{
        f\bigl(\Psi_{\tau}(\Omega_{\tau})\bigr)
        \otimes P_{\vec{0}}
        :
        f\text{ bounded}
    \right\}.
    \label{eq:FP_event_center}
\end{align}
Thus, the relevant event alternatives are determined
by the spectral projections of
$\Psi_{\tau}(\Omega_{\tau})$. The same construction can be repeated at every
subsequent time step. We therefore see that the
key reason why the event structure of the original
discrete FP model remains under control is essentially
the factorization property
(\ref{key discovery for discrete FP}), which allows
the Gram matrix to be completely absorbed into
the atomic system while leaving the $R$-fields
in an independent reference state $\phi_{\vec{0}}$.

\par
In contrast, in our adapted FP model, the analogue
of (\ref{key discovery for discrete FP}) is given
by (\ref{effective state no collapse}). The crucial
difference is the appearance of the additional
cross term $K_{mn}^{\circ}$, which depends explicitly
on the observable of the future horizon algebra.
As discussed earlier, the presence of this cross
term spoils the factorization property that makes
the original discrete FP model particularly tractable.
If this cross term was absent, the corresponding
factorized structure would be recovered, allowing
the machinery of the original FP model to be applied
in essentially the same way.

As we have seen, in the original FP model, the
determination of the event structure is reduced
to the atomic system, while the $R$-fields,
which serve as the environment, contribute no
additional event alternatives within the
relevant reference-state sector. In our adapted FP model, however, the failure
of factorization suggests that the environment,
now represented by the future horizon algebra,
may also play an important role in determining
the event structure underlying horizon-induced
decoherence. In other words, the horizon may
not merely serve as an environment that carries
away which-branch information, but may also
participate in the algebraic structure governing
the occurrence of quantum events.

\section{Discussion}
We  briefly summarize what we have done in this paper. In Section 2, we provide an overview of the Events,
Trees and Histories (ETH) approach to quantum mechanics.
In Section 3, we examine the physics of the
Fr\"ohlich--Pizzo (FP) model and demonstrate how
the essential ingredients of the ETH approach are
realized within this concrete dynamical framework.
Particular emphasis is placed on the Principle of
Diminishing Potentialities (PDP) and the intrinsic
stochasticity of quantum dynamics, as manifested
through the occurrence of quantum events. In Section 4, we revisit the central arguments of
Danielson, Satishchandran and Wald (DSW) concerning
horizon-induced decoherence. In Section 5, we propose an adaptation of the FP model
based on the local algebras of a $U(1)$ chiral current
to describe horizon-induced decoherence. Our adapted
model retains essential features of the original FP
construction while incorporating the causal structure
of a Killing horizon. We show that our adapted model
reproduces the characteristic asymptotic decoherence
behavior identified by DSW, with the reduced dynamics
of the charged particle exhibiting a form consistent
with an effective GKSL equation in the large holding-time
limit. However, unlike in the original discrete FP model,
we find that nontrivial vacuum correlations between
the interaction region and the future horizon prevent
the exact factorization of the interacting state into
a state of the charged particle and an independent
reference state of the horizon. This observation
suggests possible non-Markovian features in the
underlying dynamics, despite the apparently Markovian
behavior of the reduced system in the large holding-time
limit. We further discuss the implications of this failure
of factorization for the occurrence of quantum events
within the ETH approach. In particular, whereas the
event structure of the original FP model can be
determined from the spectral projections of the
effective atomic density matrix, the corresponding
event structure in our adapted model may also depend
nontrivially on the horizon degrees of freedom.
This suggests that the horizon may play a role not
only in inducing decoherence but also in the algebraic
structure governing the occurrence of quantum events.

\par We conclude by posing some questions and reflection from our study for one to contemplate:

\begin{itemize}

\item The first question concerns the possibility of
developing a more flexible method for determining
the structure of quantum events in our adapted FP model.
As discussed in the previous section, the ETH approach
provides an axiomatic criterion for the occurrence of
quantum events in terms of the center of the centralizer
of the physical state on the relevant algebra of
potentialities. However, applying this criterion directly
to our continuum horizon model requires a detailed
analysis of the centralizer of the interacting state,
which appears to be a rather difficult 
problem. Moreover, even if the first actual event
could be identified, determining the subsequent events
would require repeating this analysis for the conditional
state resulting from each event. It is therefore our hope that a more flexible approach
to determining the structure of quantum events can be
developed, one that remains consistent with the
fundamental principles of the ETH approach while going
beyond its present axiomatic formulation. Such an
approach might provide a more practical way of describing
the stochastic dynamics of quantum events in continuum
quantum field theory, without requiring an explicit
calculation of the center of the centralizer at every
stage of the evolution.

    \item  The second question concerns the role of \textbf{modular resonance}
in determining the structure of quantum events in the
original discrete FP model. Recall that, in our preceding
discussion, we have chosen the reference state of the
$R$-fields to be the pure product state $\phi_{\vec{0}}$.
However, one may also consider other choices, such as
a maximally mixed state or a finite-temperature Gibbs state for each cell.
It is interesting to compare these three cases and examine
how the choice of reference state affects the center of
the centralizer of the combined system. We first recall that, for a faithful product state
$\omega_A\otimes\phi$ on
$\mathcal{A}{\otimes}\mathcal{B}_{\geq\tau}$,
the corresponding modular automorphism group factorizes as $\sigma_s^{\omega_A\otimes\phi}
    =
    \sigma_s^{\omega_A}
    \otimes
    \sigma_s^{\phi}$.
One might therefore expect the centralizer of the
combined system to be simply the tensor product of
the individual centralizers. However, this conclusion
is not always correct, owing to the possibility of
modular resonance. In particular, if the atomic
system and the $R$-fields possess modular eigenoperators
with opposite modular frequencies, their tensor
product is invariant under the combined modular flow.
To see this more explicitly, suppose that
$A_{\lambda}\in\mathcal{A}$ and
$B_{-\lambda}\in\mathcal{B}_{\geq\tau}$
satisfy
\begin{align}
    \sigma_s^{\omega_A}(A_{\lambda})
    &=
    e^{i\lambda s}A_{\lambda},
    \\
    \sigma_s^{\phi}(B_{-\lambda})
    &=
    e^{-i\lambda s}B_{-\lambda}.
\end{align}
Then
\begin{align}
    \sigma_s^{\omega_A\otimes\phi}
    (A_{\lambda}\otimes B_{-\lambda})
    &=
    A_{\lambda}\otimes B_{-\lambda}.
\end{align}
Consequently,
\begin{align}
    A_{\lambda}\otimes B_{-\lambda}
    \in
    \mathcal{C}_{\omega_A\otimes\phi}
    \left(
        \mathcal{A}{\otimes}
        \mathcal{B}_{\geq\tau}
    \right).
\end{align}
Such an operator need not belong to the tensor
product of the individual centralizers. This is
the phenomenon of modular resonance.

For the pure reference state $\phi_{\vec{0}}$
considered in the original FP construction,
the state is not faithful on the full $R$-field
algebra. On its support algebra $P_{\vec{0}} \mathcal{B}_{\geq \tau} P_{\vec{0}}=\mathbb{C} P_{\vec{0}}$, however, the
modular automorphism group is trivial \footnote{Every automorphism of a one-dimensional unital algebra must be the identity automorphism.}, since
the support of a pure normal state on the
Type I representation is one-dimensional.
There are therefore no nonzero modular frequencies
arising from this support algebra that could
resonate with those of the atomic system.
This is consistent with the fact that, in the
original FP construction, the relevant event
structure can be determined from the spectral
projections of the effective atomic density
matrix, while the $R$-fields remain in an
independent pure reference state.

Now, we consider the case in which each $R$-field cell
is equipped with a maximally mixed state.
The corresponding infinite product state
is tracial, and the resulting von Neumann
algebra is naturally described by a
Type $\mathrm{II}_1$ factor. Since the
reference state is tracial, its modular
automorphism group is trivial $\sigma_s^{\phi_{\mathrm{tr}}}
    =
    \operatorname{id}$.
In this case, the centralizer of the $R$-field
algebra is the entire algebra $\mathcal{C}_{\phi_{\mathrm{tr}}}
    (\mathcal{B}_{\geq\tau})
    =
    \mathcal{B}_{\geq\tau}$. Furthermore, since the modular flow of the
tracial reference state is trivial, no
cancellation of nonzero modular frequencies
can occur between the atomic system and
the $R$-fields. For a faithful atomic state,
we therefore obtain
\begin{align}
    \mathcal{C}_{\omega_A\otimes\phi_{\mathrm{tr}}}
    \left(
        \mathcal{A}{\otimes}
        \mathcal{B}_{\geq\tau}
    \right)
    =
    \mathcal{C}_{\omega_A}(\mathcal{A})
    {\otimes}
    \mathcal{B}_{\geq\tau}.
\end{align}
Since $\mathcal{B}_{\geq\tau}$ is a factor,
the center of this centralizer is
\begin{align}
    Z_{\omega_A\otimes\phi_{\mathrm{tr}}}
    \left(
        \mathcal{A}{\otimes}
        \mathcal{B}_{\geq\tau}
    \right)
    =
    Z_{\omega_A}(\mathcal{A})
    \otimes\mathbb{1}.
\end{align}
Thus, although the centralizer is enlarged,
the center of the centralizer remains
determined by the atomic system.

The situation becomes more subtle when the $R$-fields
are equipped with a finite-temperature Gibbs state
$\phi_{\beta}$. In this case, the modular automorphism
group is generally nontrivial, and the possibility
of modular resonance must be taken into account.
Importantly, the $R$-field system in the original FP
model consists of infinitely many cells, each equipped
with the same Gibbs state. The modular flow of the
infinite tensor-product state acts as the tensor
product of the modular flows of the individual cells.
Consequently, modular frequencies arising from
different cells may cancel one another, producing
additional invariant operators in the centralizer
of the infinite $R$-field algebra. For example, if $E_{ab}^{(n)}$ denotes a matrix unit
associated with an energy transition in the $n$-th
cell, its modular flow is given by
\begin{align}
    \sigma_s^{\phi_{\beta}}
    (E_{ab}^{(n)})
    =
    e^{-i\beta(E_a-E_b)s}
    E_{ab}^{(n)}.
\end{align}
A tensor product of such operators involving finitely
many cells therefore has a modular frequency given
by the sum of their individual frequencies.
In particular, two operators with opposite modular
frequencies may combine to produce an operator
invariant under the modular flow of the infinite
system. This shows that the centralizer of the
infinite tensor-product state can be considerably
larger than the tensor product of the centralizers
of the individual cells.
For the standard construction with identical,
faithful, nontracial Gibbs states on two-level cells,
the resulting infinite tensor-product algebra is
a Type $\mathrm{III}_{\lambda}$ factor, while the
centralizer of the product state is a
Type $\mathrm{II}_1$ factor. Thus, even though
modular resonance gives rise to additional
operators in the centralizer, its center remains
trivial. A further possibility arises when we include
the ``atomic'' system. Suppose that the effective
atomic density matrix $\Omega_{\tau}$ is faithful,
with eigenvalues $p_m>0$. The modular frequencies
of the atomic system are determined by
$\log(p_m/p_n)$, while those of the infinite
$R$-field system arise from combinations of
the energy differences of its constituent cells.
Modular resonance between the atomic system and
the $R$-fields may occur whenever
\begin{align}
    \log\frac{p_m}{p_n}
    =
    \beta\sum_j(E_{a_j}-E_{b_j}).
\end{align}
In this case, additional operators may belong
to the centralizer of the combined system,
even though they do not belong to the tensor
product of the individual centralizers.  
It would therefore be interesting to understand
whether modular resonance between the atomic
system and the infinitely many $R$-field cells
can  modify the structure of quantum
events?

\item The occurrence of events, often accompanied by an element of stochasticity, is a familiar feature of the physical world, particularly in our everyday experience. Yet, in a more fundamental microscopic theory, conventional notions of spacetime, locality, and causality may no longer be applicable. If the very notion of an event also ceases to be meaningful at such a fundamental level, could the world of events that constitutes our experienced reality be understood as an emergent phenomenon? This raises a further question: can the stochastic dynamics of quantum events be recovered from a more fundamental microscopic theory through some physical mechanism or an appropriate coarse-graining procedure? If so, what would such a mechanism entail? Alternatively, might the occurrence of actual events require additional dynamical principles beyond those contained in the underlying microscopic theory?

We recognize that posing the question in this way may seem unconventional, as it touches upon issues often associated with the notion of ``strong emergence." However, we still think it is important to raise this question because \textit{`` Questioning is the piety of Thinking'' \footnote{Quoted from the article ``The Question Concerning Technology'' by Martin Heidegger.}.}

\end{itemize}

\section*{Acknowledgements}
The author has drawn much inspiration from a lecture series \cite{Froehlich2019}
by J\"urg Fr\"ohlich.
\bibliography{references}

\bibliographystyle{utphys}

\end{document}